\documentclass[onecolumn]{aastex63}

\usepackage[T1]{fontenc}
\usepackage[utf8]{inputenc}
\usepackage[letterspace=-15]{letterspace}
\usepackage{afterpage}

\received{November 2, 2020}
\accepted{January 28, 2021}
\submitjournal{Frontiers in Astronomy and Space Sciences - Exoplanets:
The Effect of Stellar Multiplicity on Exoplanetary Systems}

\shorttitle{Exoplanets in Visual Binaries}
\shortauthors{Fontanive et al.}

\graphicspath{{./}{figures/}}

\begin{document}

\title{The Census of Exoplanets in Visual Binaries: population trends from a volume-limited Gaia DR2 and literature search}

\correspondingauthor{Clémence Fontanive}
\email{clemence.fontanive@csh.unibe.ch}

\author[0000-0002-2428-9932]{Clémence Fontanive}
\affiliation{Center for Space and Habitability, University of Bern, Bern, Switzerland}

\author[0000-0001-8170-7072]{Daniella Bardalez Gagliuffi}
\affiliation{American Museum of Natural History, New York, NY, USA}

\begin{abstract}

We present results from an extensive search in the literature and \textit{Gaia}~DR2 for visual co-moving binary companions to stars hosting exoplanets and brown dwarfs within 200~pc. We found 218 planet hosts out of the 938 in our sample to be part of multiple-star systems, with 10 newly discovered binaries and 2 new tertiary stellar components. This represents an overall raw multiplicity rate of $23.2\pm1.6\,\%$ for hosts to exoplanets across all spectral types, with multi-planet systems found to have a lower stellar duplicity frequency at the 2.2-$\sigma$ level. We found that more massive hosts are more often in binary configurations, and that planet-bearing stars in multiple systems are predominantly observed to be the most massive component of stellar binaries. 
Investigations of the multiplicity of planetary systems as a function of planet mass and separation revealed that giant planets with masses above 0.1~M$_\mathrm{Jup}$ are more frequently seen in stellar binaries than small sub-Jovian planets with a 3.6-$\sigma$ difference, a trend enhanced for the most massive ($>$7~M$_\mathrm{Jup}$) short-period ($<$0.5~AU) planets and brown dwarf companions. Binarity was however found to have no significant effect on the demographics of low-mass planets ($<$0.1~M$_\mathrm{Jup}$) or warm and cool gas giants ($>$0.5~AU). While stellar companion mass appears to have no impact on planet properties, binary separation seems to be an important factor in the resulting structure of planetary systems. Stellar companions on separations $<$1000~AU can play a role in the formation or evolution of massive, close-in planets, while planets in wider binaries show similar properties to planets orbiting single stars. Finally, our analyses indicate that numerous stellar companions on separations smaller than 1--3~arcsec likely remain undiscovered to this date. Continuous efforts to complete our knowledge of stellar multiplicity on separations of tens to hundreds of AU are essential to confirm the reported trends and further our understanding of the roles played by multiplicity on exoplanets.
\end{abstract}


\section{Introduction}
\label{intro}

The architectures of stellar, sub-stellar, and planetary systems are relics of their formation and evolutionary processes. By studying the orbital parameters and configurations of hierarchical systems as an ensemble we can in principle trace back to the formation mechanisms that originated them. Planet formation is a direct consequence of star formation, yet it can be severely influenced by the presence of a stellar companion. The existence of planets in orbit around one or both components of binary systems are stringent probes of planet formation process. Radial velocity measurements estimate that $18\pm1\,\%$ of FGK stars will have a giant planet within 20~AU \citep{Cumming2008}. About $44\,\%$ of FGK stars are found in multiple systems, with $33\,\%$ in binary systems, and $11\,\%$ in higher-order architectures \citep{Raghavan2010}. Hence roughly half of potential planet hosts are in multiple-star systems, arguing that the fraction of giant planets orbiting a stellar component of a binary system is likely not negligible.

A number of campaigns have thus searched for planets in and around stellar binaries, including radial velocity programs (e.g., \citealp{Konacki2009}), transit discoveries (e.g., \citealp{Doyle2011}) and direct imaging surveys (e.g., \citealp{Asensio-Torres2018,Hagelberg2020}), leading to the detection of a number of circumstellar (orbiting one star) and circumbinary (orbiting two stars) planets.
Despite these efforts, most exoplanet searches routinely exclude stars in binary or multiple systems to avoid systematic errors in planet detection, and the first systems of this type were identified serendipitously \citep[see e.g.,][]{Patience2002,Mugrauer2006}. The distinct demographics of the first planets discovered in binary star systems hinted at the possibility that binary companions could dramatically reorient the orbital configuration of planetary systems \citep{ZuckerMazeh2002}. Approaching the question from the opposite end, numerous high-resolution imaging studies have also searched for stellar companions to known planetary systems, either to validate or refine the nature of identified planets \citep{Everett2015,Furlan2017,Hirsch2017}, or to purposely investigate the effect of stellar duplicity on planetary populations \citep{Colton2021,Horch2014,Matson2018,Wang2014}.

Dedicated studies of circumstellar planets in binary systems rapidly revealed a lack of stellar companions within 20--50~AU \citep[e.g.,][]{Bergfors2013,Kraus2016}. Close stellar companions on this separation range are generally accepted to prevent planet formation, although early examples of giant planets in $<$20~AU binaries \citep{Queloz2000,Hatzes2003} demonstrated that such systems do exist. Consistent with the observed shortfall of planets of tight binaries, theoretical models predict that the presence of a very close binary companion can truncate a protoplanetary disk \citep{Kraus2012,Pichardo2005}, hence obstructing the formation of a planet by core accretion, or ejecting the planet in unstable systems \citep{Kaib2013}. Binary companions at large separations (beyond several hundreds to thousands of AU) from planet hosts, on the other hand, have been argued to have no impact on the formation and evolution of planets \citep{DesideraBarbieri2007,WhiteGhez2001}.

Meanwhile, the effects of binary companions at intermediate (around $\sim$100--300~AU) separations are more debated. Such companions could truncate circumprimary disks by opening large gaps, hence redirecting the material to the primary stars' circumstellar disks and leaving the secondaries with no or depleted disks \citep{ArtymowiczLubow1994,BateBonnell1997}, consistent with observations of binary systems among T Tauri stars \citep{JensenAkeson2003}. Theoretical simulations also showed that perturbations from secondary stars may assist the formation and evolution of giant planets by enhancing mass accretion and orbital migration rates in circumstellar disks \citep{Kley2001}. The Kozai-Lidov mechanism \citep{Kozai1962,Lidov1962} could also play a role in the inward migration and final orbital properties of planets through secular interactions induced by an outer stellar companion on such separations. This process has indeed been invoked to explain the formation of hot Jupiters \citep{FabryckyTremaine2007,Winn2010}.

In this study, we present an overview of the current census of circumstellar exoplanets in visual binaries, with separations from tens of AU out to 20\,000~AU, within a volume limited to 200~pc. The goal of this compilation is to gather information of stellar multiplicity for a large sample of exoplanets, which will hopefully serve in future investigations, rather than to perform a detailed statistical analysis of these populations. In particular, this work extends previous such studies of exoplanets orbiting one component of a binary system to all stellar spectral types and all types of extra-solar planets and brown dwarf companions.
In Section~\ref{methods}, we describe the construction of our exoplanet sample (Section~\ref{planet_sample}), followed by a search for co-moving companions to exoplanet hosts (Section~\ref{binary_search}) in the literature and in \textit{Gaia}. Section~\ref{results} presents our results, in which we explore differences between the demographics of planets in binaries and around single stars (Section~\ref{bin_from_planet_prop}), as well as potential trends in planet properties based on binary separation and mass, for the population of planets in multiple-star systems (Section~\ref{planet_prop_from_bin}). Section~\ref{discussion} discusses the completeness of our sample and the observed effects of stellar duplicity on various exoplanetary populations. Our conclusions are presented in Section~\ref{conclusion}.

\section{Materials and Methods}
\label{methods}

We describe in this Section the construction of our studied exoplanet sample (Section~\ref{planet_sample}) and the searches performed for wide binary companions to all selected planet hosts (Section~\ref{binary_search}). In the context of this work, we consider brown dwarfs and extra-solar planets orbiting stars as a unique population of sub-stellar companions. We thus make no distinction between companions below and above the deuterium burning limit (13 M$_\mathrm{Jup}$), and will use the term sub-stellar companion to denote planetary and brown dwarf companions in general, unless otherwise specified. Similarly, double and multiple stellar systems will often be referred to as binaries throughout most of this work for conciseness. Finally, given the exoplanet-oriented approach of this study, the term host will always refer to the planet-bearing star in a system, independently of whether or not it is the higher-mass component of a multiple-star system.


\subsection{Exoplanet Compilation}
\label{planet_sample}

We gathered a sample of extra-solar planets and brown dwarfs from the NASA Exoplanet Archive\footnote{\url{https://exoplanetarchive.ipac.caltech.edu}}, the Extrasolar Planets Encyclopaedia\footnote{\url{http://exoplanet.eu}} \citep{Schneider2011}, the Exoplanet Orbit Database\footnote{\url{http://exoplanets.org}} \citep{Han2014} and the Open Exoplanet Catalogue\footnote{\url{http://www.openexoplanetcatalogue.com}}. The data from these libraries were collected on June 23, 2020, and cross-matched to identify all systems with at least one planet or brown dwarf companion reported as confirmed in at least one of these databases. We gathered from these catalogs all available information about the sub-stellar companions and stellar hosts, and only kept systems with robust companion mass (or minimum mass) and semi-major axis measurements. We imposed a cut of 0.1~M$_\odot$ on the minimum mass of the host, based on primary masses supplied in the considered databases, in order to focus our study on stellar hosts only. We also removed all circumbinary (P-type) systems, orbiting both stars from a binary system, as our study concentrates on circumstellar (S-type) planets and brown dwarfs, found around a single component of a binary system.

We cross-matched the resulting sample with the \textit{Gaia} Data Release~2 (DR2; \citealp{Gaia2016,Gaia2018}) catalog, obtaining positions, parallaxes, proper motions, \textit{Gaia} magnitudes and effective temperatures for all hosts found in \textit{Gaia}~DR2. Astrometric information was taken from the SIMBAD Astronomical Database \citep{Wenger2000} for the few targets that are not part of \textit{Gaia}~DR2 or do not have full astrometric solutions from the \textit{Gaia} mission. Based on the obtained stellar parallaxes, we restricted our sample to systems with parallax measurements larger than 5~mas, corresponding to a maximum distance of 200~pc for our volume-limited investigation. This cut allows us to focus on relatively nearby stars, thus limiting the range of probed inner working angles around different targets when searching for stellar companions, while keeping a sufficiently large sample for a statistically significant study.

\begin{figure}
\begin{center}
\includegraphics[width=0.8\textwidth]{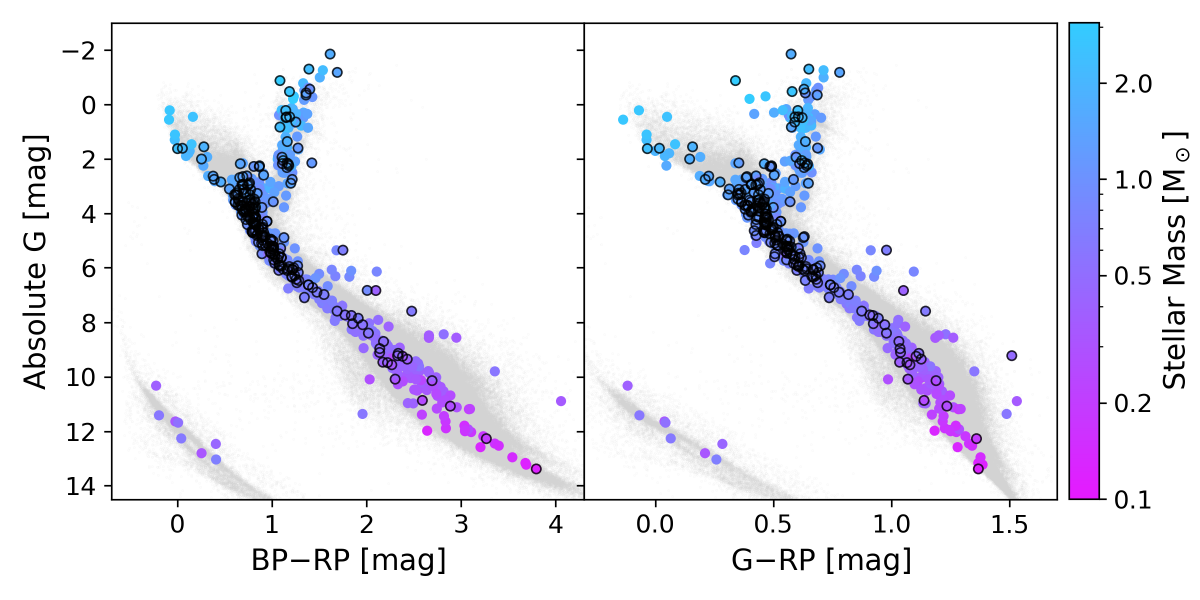}
\end{center}
\caption{\textit{Gaia} color-magnitude diagrams of planet hosts stars, showing absolute \textit{G} magnitudes against \textit{BP}-\textit{RP} colors (left) and \textit{G}-\textit{RP} (right). Symbols plotted with black rings represent planet hosts found to be part of multiple-star systems. The colorbar indicates the mass of each planet host, using a logarithmic scale. The gray background population shows the 200-pc volume-limited cleaned sample from \textit{Gaia}~DR2.}
\label{f:gaia_cmd_hosts}
\end{figure}

The final sample consists of 938 host stars, harboring a total of 1316 exoplanets and brown dwarfs, and contains 693 single-planet systems and 245 multi-planetary systems. Stellar hosts have masses ranging from 0.1 to 3.09~M$_{\odot}$, with a median of 0.95~M$_{\odot}$.
Most primaries are along the main sequence, covering spectral types from B to M, with 171 giants or sub-giants and 8 white dwarfs.
We show in Figure~\ref{f:gaia_cmd_hosts} the \textit{Gaia} Hertzsprung-Russell diagrams for all primaries in our sample with \textit{Gaia} DR2 parallaxes and \textit{G}, \textit{BP}, and \textit{RP} magnitudes (926 stars), with the color scale indicating the host mass. Tables for the final samples are provided as supplementary material and are available online, with separate tables for the stellar hosts and sub-stellar companions.

\subsection{Binary Search}
\label{binary_search}

In this work, we focus on co-moving visual binaries or higher-order hierarchical stellar systems, that is, systems with two or more stars confirmed to be moving together in the sky. The co-moving nature of two gravitationally-bound objects can be determined in two ways. The first approach is via proper motion (and parallax) measurements, where the components of a multiple system will show astrometric parameters consistent with one another. The second method consists in comparing images taken over a sufficiently-long time baseline to demonstrate that two sources show the same displacement over time compared to fixed background objects. Both approaches require the two (or more) stars to be spatially resolved in imaging observations, and such visual binaries are thus typically widely-separated systems. We place an outer limit of 20\,000~AU on the projected separation in our search for multiple systems.

In the following sections, we describe the searches we performed for wide, co-moving visual companions to all stellar hosts from our gathered sample of exoplanetary systems. The full compilation of binary systems is provided as online supplementary material.

\subsubsection{Binaries in Surveys and the Literature}
\label{lit_search}

The catalogs used to compile our studied sample contain some information about stellar binarity. We complemented the multiplicity data from these databases with the Catalogue of Exoplanets in Binary Star Systems\footnote{\url{https://www.univie.ac.at/adg/schwarz/multiple.html}} \citep{Schwarz2016}. We added to this all systems from published surveys searching for visual stellar companions to circumprimary planetary systems \citep{Adams2012,Adams2013,Bergfors2013,Bohn2020,Coker2018,Daemgen2009,Deacon2016,DietrichGinski2018,Eggenberger2007,Eggenberger2011,Faedi2013,Fontanive2019,Furlan2017,Ginski2012,Ginski2016,Ginski2020,Kraus2016,Lillo-Box2012,Lodieu2014,LuhmanJayawardhana2002,Moutou2017,Mugrauer2006,Mugrauer2007b,Mugrauer2007,MugrauerGinski2015,MugrauerNeuhauser2009,Mugrauer2019,Ngo2016,Ngo2017,Patience2002,Raghavan2006,Southworth2020,Udry2004,Wang2014,Wollert2015,Ziegler2018} or reviews of planets in binaries \citep{BonavitaDesidera2007,BonavitaDesidera2020,DesideraBarbieri2007,EggenbergerUdry2007,Eggenberger2010,Roell2012,ThebaultHaghighipour2015} that we could find, and finally any other serendipitous discovery we were aware of that may have been missing from the above compilations.

In parallel, we cross-matched our host star sample with large-scale catalogs of stellar multiplicity like the Washington Double Star Catalog (WDS; \citealp{Mason2001}), the Catalog of Components of Double \& Multiple stars (CCDM; \citealp{Dommanget2002}), the Tycho Double Star Catalogue (TDSC; \citealp{Fabricius2001}) and the Updated Multiple Star Catalog (MSC; \citealp{Tokovinin2018}), as well as surveys for wide stellar binaries conducted with direct imaging \citep{Deacon2014,Janson2012,Janson2014,Janson2017,Raghavan2010,Tokovinin2012,Tokovinin2014,Tokovinin2014b,Ward-Duong2015,Winters2019}. 

Each reported multiple system was then checked individually in the literature to ensure the S-type nature of the planets and brown dwarfs, and confirm that the binary or multiple system was indeed visual (with resolved components), and astrometrically confirmed to be co-moving, either via consistent relative astrometry in multi-epoch observations or through similar kinematics (as opposed to optical binaries with a probabilistic bound nature from the chance of alignment). 
A total of 184 stars in our sample were mentioned in the considered surveys to have at least one companion satisfying these criteria (excluding the recent \textit{Gaia} search performed by \citealp{Mugrauer2019}; see Section~\ref{gaia_search}).
For all identified systems, we gathered, when available, binary separations, companion masses, and companion spectral types.

\subsubsection{Companions in Gaia DR2}
\label{gaia_search}

To complement the literature search performed above, we searched for bright companions in the \textit{Gaia}~DR2 catalog to all stars in our compilation. Using the collected positions, proper motions and parallaxes for the stellar hosts, we searched for \textit{Gaia} sources within angular distances corresponding to separations of 20\,000~AU from our primaries, and displaying consistent kinematics. Following the approach from \citet{Fontanive2019}, we used thresholds of $20\,\%$ disparity in parallax, and offsets of $<20\,\%$ of the total proper motion in one direction and $<50\,\%$ in the other coordinate. These cuts allow to account for the fact the short-term astrometric measurements from \textit{Gaia}~DR2 may capture the reflex motion of binary systems, or may have spurious solutions for unresolved binaries (see \citealp{Fontanive2019} for details). 

For systems part of young moving groups, other members of the same association may appear nearby in the sky and display similar proper motions and parallaxes, consistent with the average moving group kinematics. To avoid the inclusion of unassociated close-by group members in our binary list, we checked that no more than one other astrometric match was found on angular separations up to 20 times the identified binary radius. We consider that a co-moving source within 20\,000~AU projected separation is statistically unlikely to be an unrelated member of the same group if no other members are found within a 400-fold sky area. We thus regard such sources as bonafide bound companions for the purpose of this study. When one other match was found, we applied the same procedure centered on this outer source, with the same search radius, to establish whether other group members were found nearby, in which case all sources were taken to be unrelated moving group members. If no additional sources with consistent kinematics were found, we considered the outer source to be the tertiary component of a triple system.
Finally, we checked that identified binary companions were different from the sub-stellar companions in our exoplanet list, as some young and bright brown dwarfs discovered with direct imaging on wide separations may be detected at the low-mass end of the \textit{Gaia}~DR2 completeness \citep{Reyle2018}.

This analysis yielded 175 companions around 172 hosts stars.
For all identified systems, we measured the binary separation from the respective \textit{Gaia}~DR2 positions of co-moving components, and collected \textit{Gaia} photometry for the companions. The majority (139) of identified \textit{Gaia} companions were already known from the literature (excluding findings from \citealp{Mugrauer2019}) and were included in our compilation from Section~\ref{lit_search}. Based on our literature findings, 19 of the detected \textit{Gaia} companions were in fact tight binaries themselves, unresolved in \textit{Gaia}.
\citet{Mugrauer2019} recently performed a very similar search for wide companions to exoplanet host stars in \textit{Gaia}~DR2, which presents a useful comparison survey to validate our approach. We found that all binaries reported in that work and present in our sample list (121 systems) were also retrieved in our \textit{Gaia} analysis. From these, 23 systems and a tertiary companion to the WASP-11 system (unresolved in \textit{Gaia}) were never reported prior to that study. We however retrieved 51 additional \textit{Gaia} systems missing from the \citet{Mugrauer2019} compilation, which is likely due to different target samples between that study and ours. 
Finally, 10 of our identified \textit{Gaia} co-moving systems were not found to have been previously reported in the literature (up to early September 2020): CoRoT-7, HD~13167, HD~23472, HIP~73990, K2-228, L2~Pup, TOI-132, WASP-189, WASP-29, WASP-59. In addition, new tertiary components were discovered around HIP~65A and V~1298~Tau. These companions are presented in Table~\ref{t:new_systems}, with additional information about the systems provided within the full catalogs available online.

\begin{table}
    \centering
    \caption{New stellar companions to planet host stars identified in this work using the \textit{Gaia}~DR2 catalog. The new system components are marked in bold in the System column, which is only indicative of the overall system architecture, and do not represent proposed naming conventions. Following the approach in the main tables, component A systematically denotes the planet host irrespectively of the relative component masses, unless already named differently in the literature. Indices 1 in the stellar mass and spectral type columns refer to the planet hosts, and indices 2 refer to the considered stellar companions. \textit{This is a highlight of full tables available as supplementary material, which provide additional information about these systems, together with the rest of our compilation.}}
    \renewcommand{\arraystretch}{1.3}
    \begin{tabular}{ l c c c c c c c c }
    \\
    \hline\hline
        System & Parallax & Separation & Separation & SpT$_1$ & Mass$_1$ & SpT$_2$ & Mass$_2$ \\
         & [mas] & [arcsec] & [AU] & & [M$_\odot$] & & [M$_\odot$] \\
    \hline
        \multicolumn{2}{l}{New binary companions} \\
    \hline
        CoRoT-7 Abc + \textbf{B} & 6.23 & 75.7 & 12160 & K0V & 0.93 & M4 & 0.23 \\
        HD 13167 Ab + \textbf{B} & 6.69 & 20.1 & 3001 & G3V &	1.35 & M4 & 0.21 \\
        HD 23472 Abc + \textbf{B} & 25.59 & 9.6 & 374 & K3.5V & 0.75 & M6 & 0.14 \\
        HIP 73990 Abc + \textbf{B} & 9.03 & 47.3 & 5234 & F2IV & 1.72 & M2 & 0.50 \\
        K2-228 Ab + \textbf{B} & 7.71 & 22.6 & 2933 & K6 & 0.71	& G3 & 1.08 \\
        L2 Pup Ab + \textbf{B} & 15.61 & 32.8 & 1998 & M5IIIe	& 0.66 & M5 & 0.20 \\
        TOI-132 Ab + \textbf{B} & 6.08 & 19.6 & 3231 & G8V & 0.97 & M4 & 0.17 \\
        WASP-189 Ab + \textbf{B} & 10.00	& 9.4 & 942 &	A4/5IV/V & 1.89 & M2 & 0.45 \\
        WASP-29 Ab + \textbf{B} & 11.39 & 125.2 & 10994 &	K4V	& 0.82 & M3 & 0.38 \\
        WASP-59 Ab + \textbf{B} & 8.60 & 81.8 & 9512 & K5V	& 0.72 & K7 & 0.62 \\
    \hline
        \multicolumn{2}{l}{New tertiary companions} \\
    \hline
        HIP 65 (Ab + B) + \textbf{C} & 16.16 & 73.6 & 4557 &	K4V & 0.78 & M7.5 & 0.11 \\
        V 1298 Tau Ab-e + (B\textbf{C}) & 9.21 & 117.0 & 14795 &	K1 & 1.10 & K7 & 0.66 \\
    \hline
    \end{tabular}
    \label{t:new_systems}
\end{table}

For all new \textit{Gaia} systems, we checked whether the wide stellar companions were known (seemingly single) objects in SIMBAD, and gathered additional mass and spectral type information from the literature for these components when available. In addition, for all components known from the literature but missing from our \textit{Gaia} binary list, we searched for these wide companions in \textit{Gaia}~DR2 in case these stars were in the catalog but with no astrometric solutions, and thus not detectable as co-moving sources in our \textit{Gaia} search. From these, 14 companions were recovered without \textit{Gaia}~DR2 astrometry, and available \textit{Gaia} magnitudes and relative positions of components were added to our catalog for these additional companions.

\subsubsection{Properties of Stellar Companions}\label{TIC}

As a number of literature systems (mostly from WDS) and \textit{Gaia} binaries had no existing stellar classification or measured mass, we estimated these characteristics for all identified \textit{Gaia} components based on their positions in the \textit{Gaia} color-magnitude diagrams. Hertzsprung-Russell diagrams were made for all binary companions with measured magnitudes in the \textit{G}, \textit{BP} and \textit{RP} bands, using the sources' parallaxes if available, and the astrometry from the associated planet hosts otherwise.

From these, 6 sources populated the white dwarf part of the parameter space, and were all found in the \textit{Gaia}~DR2 white dwarf study by \citet{GentileFusillo2019}. The white dwarf classifications and masses were thus taken from this work for these companions.

All other companions appeared to fall along the main sequence. For these systems, we used the \textit{TESS} Input Catalog (TIC; \citealp{Stassun2018}) to map the parameter space of the \textit{Gaia} color-magnitude diagrams to stellar masses and spectral types, based on TIC stellar masses and queried SIMBAD spectral types for all sources from the catalog out to 200~pc. As clear and continuous trends in mass and spectral type were seen along the \textit{Gaia} main sequences of the TIC sample (similar to our host sample in Figure~\ref{f:gaia_cmd_hosts}), the masses and spectral types of wide companions could be inferred directly based on their location along these \textit{Gaia} main sequences. Quantities were interpolated using the mean mass and spectral type from the TIC sample in a box of size 0.2~mag centered on the companion's absolute magnitude and color, provided that at least 10 sources were found in that box. For each detected \textit{Gaia} companion, masses (rounded to 0.01~M$_\odot$) and spectral types (to 1 sub-type) were obtained from the TIC \textit{BP}--\textit{RP} and \textit{G}--\textit{RP} parameter spaces, and averaged for more robust final values. For sources characterized in this way from their colors and magnitudes, the average offset and scatter between literature values and our \textit{Gaia}-derived estimates was $+0.3\pm1.5$ sub-type in spectral type, and $-0.01\pm0.05$~M$_\odot$ in mass (removing known unresolved sources). We also validated this method by applying it to our main sequence host star sample, and observed comparably negligible offsets to values collected in the planet-host catalog.

For companions that fell outside the TIC main sequence due to unusual \textit{Gaia} colors (14 objects), or for sources with no \textit{BP} and \textit{RP} magnitudes (16 objects), we used the median intersection of the absolute \textit{G} magnitude with the TIC main sequences instead, assuming that these objects were single, main sequence stars, similar to the approach followed in the recent \textit{Gaia} study by \citet{Mugrauer2019}. For companions classified from their absolute magnitudes alone, the average scatter was $-0.4\pm2.4$ sub-types in spectral type and $-0.2\pm0.07$~M$_\odot$ in mass for sources truly on the main sequence, with very similar results in mass to \citet{Mugrauer2019} for overlapping systems. Larger offsets were seen for known white dwarf companions with no \textit{Gaia} colors, which were hence assimilated to M dwarfs on the main sequence based on their absolute \textit{G}-band magnitudes.

Based on these results, we consider that our \textit{Gaia}-inferred quantities are robust measurements for main sequence components. We adopt these as final values when no previous mass and spectral type estimates were available for the retrieved companions, and use existing literature estimates otherwise. The literature, \textit{Gaia}, and final adopted values are all reported in our tables.

\afterpage{
\begin{figure}
\begin{center}
\includegraphics[width=\textwidth]{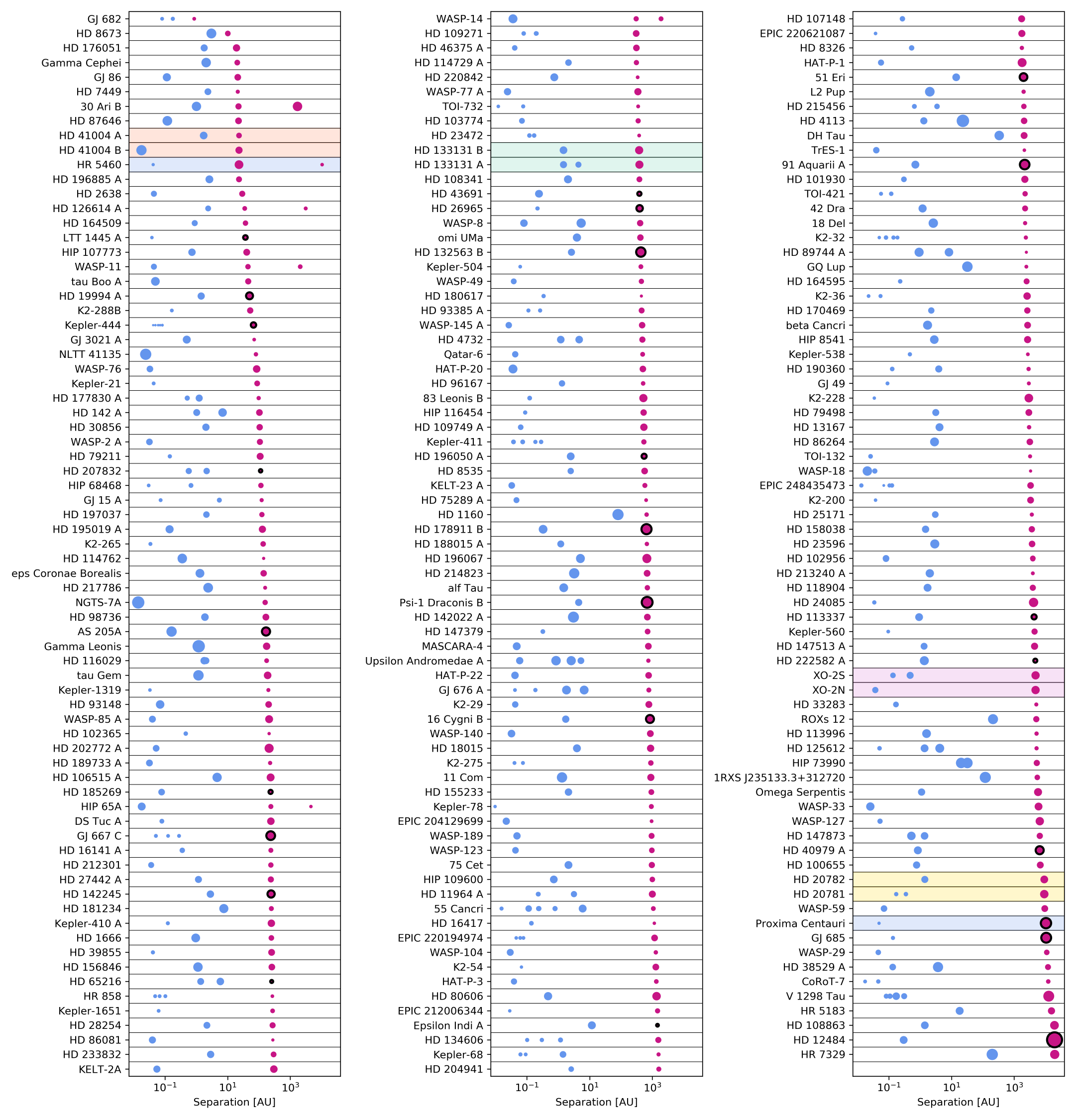}
\end{center}
\caption{Architectures of all exoplanetary systems identified to be in multiple-star configurations. The blue circles represent the inner brown dwarfs and planets, with symbol sizes proportional to their masses cubed. Magenta symbols show the positions of all confirmed wide stellar companions, with radii proportional to the mass of these outer companions. Stellar companions marked with black circles are themselves tight binaries, with the symbol sizes based on the combined mass of the two components. The 5 host stars in multiple systems containing two planet-bearing stars are color-coded accordingly. We note that separations for inner sub-stellar companions correspond are semi-major axes, while observed projected separations are displayed for the wide stellar companions.}
\label{f:architectures}
\end{figure}
\clearpage}

\section{Results}
\label{results}

Many surveys looking for extra-solar planets, in particular with the radial velocity method, are affected by or biased against binaries with separation $\leq2$--6~arcsec, excluding known multiple systems in target selection processes (see e.g., \citealp{Eggenberger2010,Ngo2017}). As a result, measurements of multiplicity rates for exoplanetary systems are particularly challenging, as these selection biases are not trivial to quantify and correct for (see e.g., \citealp{MoeKratter2019}). This typically means that studies like ours, investigating the binarity of planetary systems discovered partly by such surveys, cannot be used to derive the true frequency of planets in binaries, nor to probe the existence of planets in very tight binaries. With this in mind, the goal of this work is thus to provide an overview the current census of sub-stellar companions in wide visual binaries, rather than to achieve robust statistical results, and we will therefore not attempt to account for these biases here.

Our studied sample of planetary systems was nonetheless compiled independently from the binary nature (known or unknown) of the systems. The gathered compilation should thus not be biased toward or against the existence of binary star systems beyond the intrinsic biases from exoplanet detection campaigns. Our \textit{Gaia} search for wide companions is also homogeneous across the host star sample, limited only in inner working angle by the distance to each star, and by the inherent completeness of \textit{Gaia}~DR2. Our ability to recover stellar companions in \textit{Gaia} is therefore, in principle, independent of the architecture of the planetary systems themselves. Similarly, the existing literature surveys considered spanned a large range of planet host stars and probed various distinct planetary populations. We thus consider that while strong biases remain in our binary list, which should be taken into account for detailed statistics and the derivation of absolute occurrence rates, our compilation does not strongly discriminate between different types of sub-stellar companions (i.e., planet or brown dwarf masses, separations or detection methods) in the potential to detect wide visual companions. Our compilation can hence be used to search for raw trends within the obtained sample of binaries and highlight potential correlations between multiplicity and the properties of planetary and sub-stellar companions.

\subsection{Overall Compilation}

\begin{figure}
\begin{center}
\includegraphics[width=0.55\textwidth]{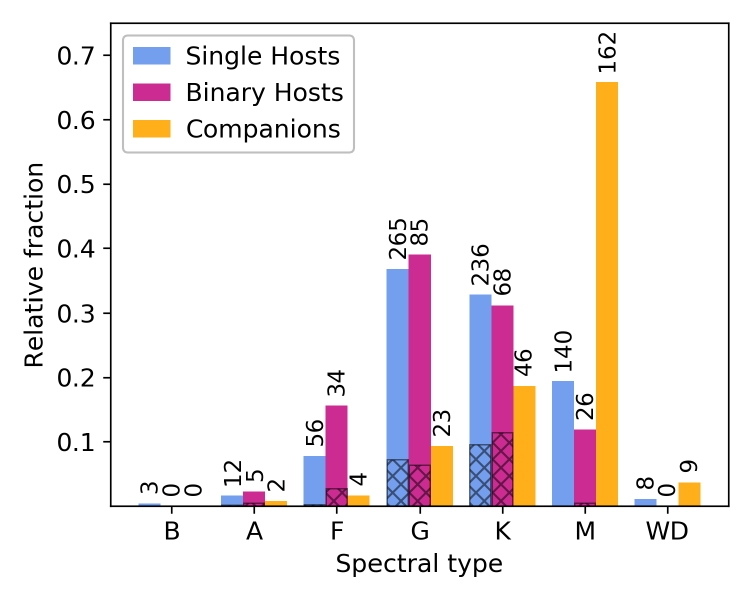}
\end{center}
\caption{Distribution of spectral types from B through M, plus white dwarfs, for single-star planet hosts (blue), multiple-star planet hosts (magenta) and stellar companions (yellow). Hatched sections of the plotted bars represents giants and sub-giants, with the remaining systems being on the main sequence. Each color-coded histogram is independently normalized so that the sum of the bars within each individual group adds up to 1.}
\label{f:SpT_histogram}
\end{figure}

From the compilation gathered in Section~\ref{lit_search}, combining an extensive literature search and a \textit{Gaia}~DR2 investigation, 218 planet hosts were found to have at least one visual co-moving stellar companion: 186 host stars were found to be in binary systems, and 32 host stars in higher-order hierarchical systems. From these, 4 binaries and 1 triple system are composed of 2 planet-hosting stars, organizing the 218 planet hosts into 213 unique multiple systems. The architecture of each planet-bearing multiple-star system is presented in Figure~\ref{f:architectures}, which illustrates the relative separations and masses of sub-stellar and stellar companions within each system. Figure~\ref{f:SpT_histogram} presents the distribution of spectral types among the planet hosts stars, showing relative numbers of single stars and planet hosts in multiple systems for each spectral type, and compared to the sample of detected stellar companions. Companion masses range from 2.37~M$_\odot$ down to the hydrogen-burning limit ($\sim$0.07~M$_\odot$). Binary projected separations extend from 0.85~AU (GJ~682) out to our 20\,000~AU search limit, with a median value of 678~AU. A total of 19 binaries were found in the range 10--50~AU, and 27 systems were identified on separations shorter than 100~AU.

Table~\ref{t:results} summarizes the numbers and raw fractions of sub-stellar companions in single and multiple-star systems, where binaries and triples are counted similarly as hierarchical multiple systems. We emphasize once again that no completeness or selection bias corrections have been performed, and the quoted numbers simply provide an overview of the collected catalogs.
From the 1316 exoplanets and brown dwarfs in our compilation, 286 were found to be around one component of a multiple-star system ($21.7\pm1.3\,\%$). In terms of individual planetary systems, 218 out of 938 planet host stars ($23.2\pm1.6\,\%$) are part of multiple-star systems. Interestingly, a marginally higher fraction (2.2-$\sigma$) of single-planet systems are in hierarchical stellar systems ($25.1\pm1.9\,\%$) compared to multi-planet systems ($18.0\pm2.7\,\%$).

\begin{table}
    \centering
    \caption{Summary of results, providing the number of single and multiple (binary or higher-order) systems hosting various planetary sub-populations. Raw occurrence rates are given in parentheses with uncertainties computed as Poisson noise.}
    \renewcommand{\arraystretch}{1.4}
    \setlength{\tabcolsep}{12pt}
    \begin{tabular}{ l c c c }
    \\
    \hline \hline
        Planetary population & Total & Single-star systems & Multiple-star systems \\ 
    \hline
    All Planets & 1316 & 1030~~($78.3\pm2.4\,\%$) & 286~~~($21.7\pm1.3\,\%$) \\
    \hline
    All Planetary Systems & 938	& 720~~~($76.8\pm2.9\,\%$) & 218~~~($23.2\pm1.6\,\%$) \\
    Single-Planet Systems & 693 & 519~~~($74.9\pm3.3\,\%$) & 174~~~($25.1\pm1.9\,\%$) \\
    Multi-Planet Systems & 245 & 201~~~($82.0\pm5.8\,\%$) & 44~~~~($18.0\pm2.7\,\%$) \\
    \hline
    M$_\mathrm{pl}<0.1$ M$_\mathrm{Jup}$ & 554 & 462~~~($83.4\pm3.9\,\%$) & 92~~~~($16.6\pm1.7\,\%$) \\
    M$_\mathrm{pl}=0.1-7$ M$_\mathrm{Jup}$ & 597 & 444~~~($74.4\pm3.5\,\%$) & 153~~~($25.6\pm2.1\,\%$) \\
    M$_\mathrm{pl}>7$ M$_\mathrm{Jup}$ & 165 & 124~~~($75.2\pm6.7\,\%$) & 41~~~~($24.8\pm3.9\,\%$) \\
    \hline
    a$_\mathrm{pl}<0.5$ AU & 766 & 603~~~($78.7\pm3.2\,\%$) & 163~~~($21.3\pm1.7\,\%$) \\
    a$_\mathrm{pl}=0.5-10$ AU & 476 & 365~~~($76.7\pm4.0\,\%$) & 111~~~($23.3\pm2.2\,\%$) \\
    a$_\mathrm{pl}>10$ AU & 74 & 62~~~($83.8\pm10.6\,\%$) & 12~~~~($16.2\pm4.7\,\%$) \\
    \hline
    M$_\mathrm{pl}\geq0.1$ M$_\mathrm{Jup}$, a$_\mathrm{pl}\leq10$ AU & 688 & 506~~~($73.5\pm3.3\,\%$) & 182~~~($26.5\pm2.0\,\%$) \\
    M$_\mathrm{pl}\geq0.1$ M$_\mathrm{Jup}$, a$_\mathrm{pl}\leq0.5$ AU & 236 & 164~~~($69.5\pm5.4\,\%$) & 72~~~~($30.5\pm3.6\,\%$) \\
    M$_\mathrm{pl}\geq7$ M$_\mathrm{Jup}$, a$_\mathrm{pl}\leq10$ AU & 106 & 73~~~~($68.9\pm8.1\,\%$) & 33~~~~($31.1\pm5.4\,\%$) \\
    M$_\mathrm{pl}\geq7$ M$_\mathrm{Jup}$, a$_\mathrm{pl}\leq0.5$ AU & 28 & 19~~~($66.9\pm15.6\,\%$) & 9~~~~($32.1\pm10.7\,\%$) \\
    \hline
    \\
    \end{tabular}
    \label{t:results}
\end{table}

\subsection{Multiplicity as a Function of Planet Properties}
\label{bin_from_planet_prop}

\begin{figure}
\begin{center}
\includegraphics[width=0.7\textwidth]{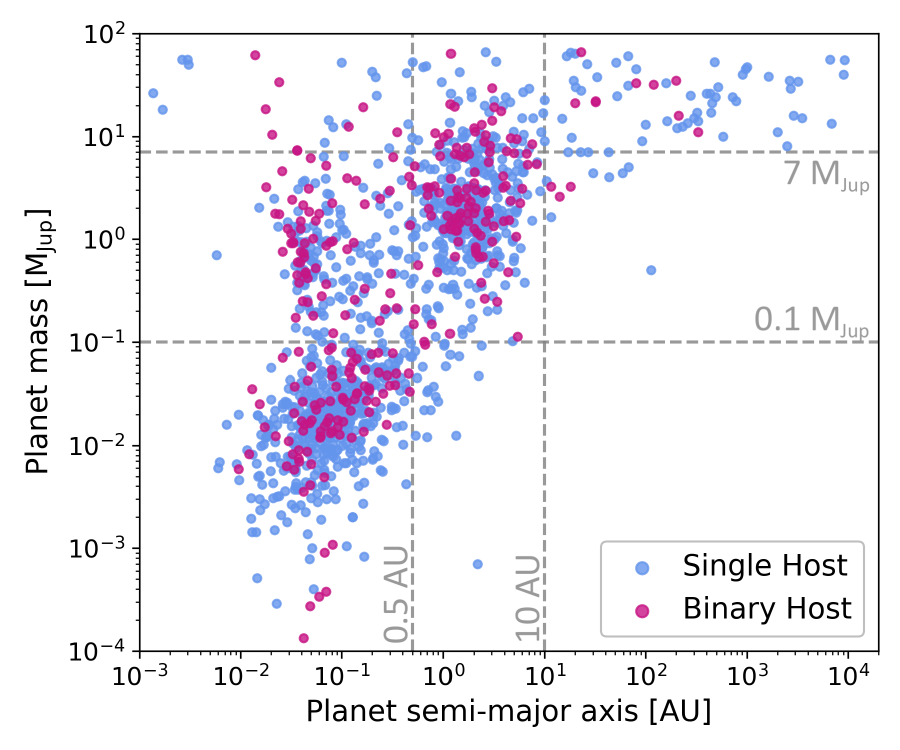}
\end{center}
\caption{Planet mass against semi-major axis for all sub-stellar companions in our exoplanet compilation. Planets identified to be part of multiple-star systems are shown in magenta, while planets orbiting single stars are plotted in blue. The dashed lines divide the parameter space into several bins detailed in the text.}
\label{f:planet_mass-sma}
\end{figure}

In this section, we explore the multiplicity of our planet host star sample as a function of planetary mass and separation. Unfortunately, other orbital elements (eccentricity, inclination) are not available for the full exoplanet sample. Investigations involving these parameters would thus be limited to planetary systems detected with specific methods and are not explored here.
In Figure~\ref{f:planet_mass-sma}, we show the masses and semi-major axes of all planet and brown dwarfs in our compilation, with systems found to be in visual stellar binaries marked in magenta, and apparently single stars in blue.

Some previously-known trends associated with specific sub-populations of planets are visually apparent in Figure~\ref{f:planet_mass-sma}. One such feature is the lack of blue scatter points (single-star systems) for sub-stellar companions with semi-major axis in the range $\sim$0.01--0.10~AU (orbital periods of $\sim$0.5--10~days around a Sun-like star) and masses larger than $\sim$3~M$_\mathrm{Jup}$. This part of the parameter space, representing massive hot Jupiters and brown dwarfs, is entirely filled with multiple-stars systems (magenta scatter points), consistent with early observations that these planets and brown dwarfs are almost exclusively observed in binary stars \citep{ZuckerMazeh2002}.
A second notable attribute from Figure~\ref{f:planet_mass-sma} is the small group of brown dwarfs with even shorter orbital separations ($<$0.01~AU) identified around single stars (top left corner). These sub-stellar companions are all found to orbit white dwarfs, and correspond to most white dwarf hosts from our compilation. Such extreme systems are thought to result from the considerable mass loss stars undergo as they become white dwarfs. This post-main sequence process drastically changes the star-planet mass ratios, thus altering the dynamics and stability of brown dwarfs and planets, in particular in multi-planet systems (e.g., \citealp{Maldonado2020}).

In order to investigate the effect of stellar multiplicity as a function of sub-stellar companion mass and separation, we divide the planetary parameter space into 3 bins in semi-major axis (a$_\mathrm{pl}$) and 3 bins in mass (M$_\mathrm{pl}$), delimited by the dashed lines in Figure~\ref{f:planet_mass-sma}. We chose arbitrary limits of 0.5 and 10~AU in semi-major axis, and 0.1 and 7~M$_\mathrm{Jup}$ in mass. The boundary at 0.5~AU corresponds to the observed dearth between two distinct peaks in the distribution of exoplanet orbital periods, representing the pile-up of hot planets, and the bulk population near the snow line ($\sim$1--3~AU), respectively \citep{Udry2003}. The 10-AU threshold corresponds roughly to the outer detection limit for the radial velocity method, and only massive, directly-imaged companions are typically identified beyond 10~AU. The 0.1-M$_\mathrm{Jup}$ mass bound was adopted as the lower limit for the mass of Jovian planets \citep{Mordasini2018}, while 7~M$_\mathrm{Jup}$ was taken as the median transition between core accretion and gravitational instability giant planets (4--10~M$_\mathrm{Jup}$; \citealp{Schlaufman2018}), a limit also advocated by \citet{MoeKratter2019} (see also \citealp{Santos2017}).

Table~\ref{t:results} reports the relative numbers of sub-stellar companions in single and binary systems in each planetary semi-major axis and mass bin. Stars harboring low-mass, sub-Jovian planets (M$_\mathrm{pl}<0.1$~M$_\mathrm{Jup}$) appear to have a substantially lower stellar binary rate, with $16.6\pm1.7\,\%$ of such planets being found in multiple-star systems. This compares to $25.5\pm1.8\,\%$ for higher-mass planets and brown dwarfs, with a 3.6-$\sigma$ difference in raw multiplicity frequency between planetary and sub-stellar companions below and above 0.1~M$_\mathrm{Jup}$. A similar trend is seen with planet orbital distance, where sub-stellar companions with a$_\mathrm{pl}>10$-AU are less frequently found in stellar binaries, although the smaller number of such planetary companions reduces the significance of this tendency. This effect is most likely the result of an enhanced bias against the existence of wide binaries within 20\,000~AU for systems with sub-stellar companions large orbital distances. Indeed, the presence of a planet or brown dwarf prevents the possibility of finding a binary companion on comparable or marginally larger separations than the sub-stellar companion semi-major axis, and binaries with separations of hundreds to thousands of AU are thus dynamically impossible for a sizable fraction of these planetary systems. Given these results, we also report values at the end of Table~\ref{t:results} focusing exclusively on the close-in (a$_\mathrm{pl}<10$~AU and $<$0.5~AU) giant planet and brown dwarf populations. While the lower number of systems associated with these subsets decreases again the significance of observed trends, raw multiplicity rates seem to increase up to around $30\,\%$ for the very shortest-separation and most massive sub-stellar companions. We also note that the vast majority of sub-Jovian planets, with masses below $0.1$~M$_\mathrm{Jup}$, are found in orbits with semi-major axes shorter than 0.5~AU.

\begin{figure}
\begin{center}
\includegraphics[width=\textwidth]{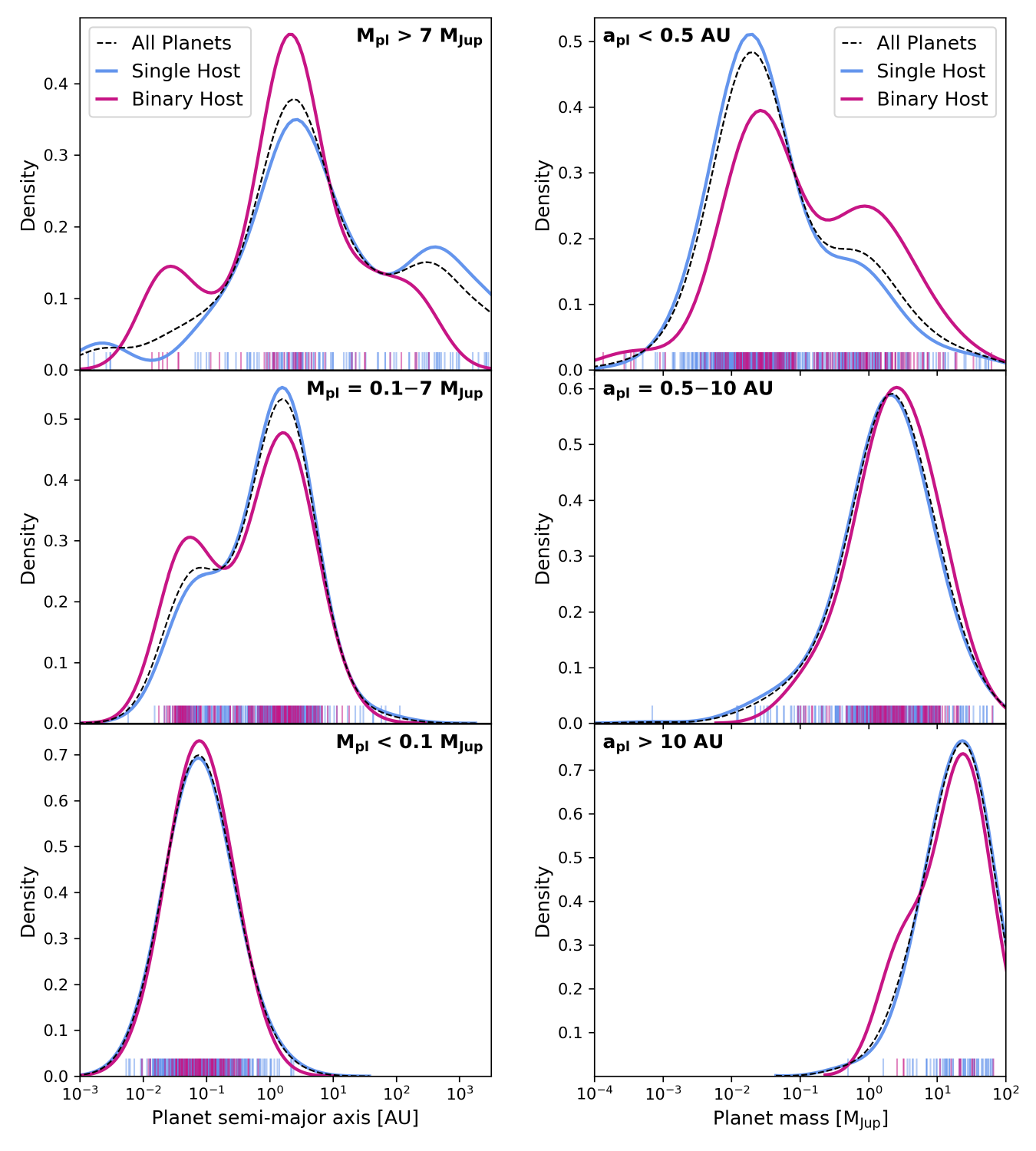}
\end{center}
\vspace{-0.3cm}
\hspace{0.25\textwidth}\textbf{(A)\hfill(B)}\hspace{0.25\textwidth}
\caption{KDEs of planet properties comparing planets in binaries (magenta) and planets around single stars (magenta), with the full planetary population shown in the dotted black lines. Panel A shows the distribution of planetary semi-major axis, divided between massive giant planets and brown dwarfs (top), lower-mass giants (middle) and sub-Jovian planets (bottom), following the cuts in parameter space shown in Figure~\ref{f:planet_mass-sma}. Panel B shows the distribution of planetary mass for close-in planets (top), intermediate-separation planets (middle) and wide-orbit giant planets (bottom).}
\label{f:kde_planet_prop}
\end{figure}

To better understand these tendencies and the effect of multiplicity with planet and brown dwarf properties, we explore the distributions of sub-stellar companions around single and binary stars in the various mass and separation bins considered. In Figure~\ref{f:kde_planet_prop}, we show kernel density estimates (KDE) of the distributions of planet semi-major axis (left panels) and mass (right panel), for the different regions of the parameter space described above. Planets and brown dwarfs in single-star systems are shown in blue, and those in hierarchical stellar systems in magenta. We use KDE bandwidths of 0.3 in all cases, and consider that such estimates of the probability density functions should provide good insights into potential underlying trends.

In terms of planet semi-major axis (Figure~\ref{f:kde_planet_prop}A), multiplicity appears to have no effect on the orbital separation of sub-Jovian planets (M$_\mathrm{pl}<0.1$~M$_\mathrm{Jup}$), illustrated by the perfectly consistent distributions for single and binary hosts in the bottom panel, both showing the same narrow peak in the semi-major axis distribution around 0.1~AU. As we enter the giant planet regime (M$_\mathrm{pl}=0.1$--7~M$_\mathrm{Jup}$; middle panel), the bulk of the planetary population shifts to separations of 1--3~AU, with a secondary peak at tighter separations (a$_\mathrm{pl}<0.1$~AU). The relative density of planets in this secondary sub-population seems to be marginally higher for binary-star systems. Looking at the most massive giant planets and brown dwarf companions (M$_\mathrm{pl}>7$~M$_\mathrm{Jup}$; top panel), a number of new features emerge in the plotted KDEs. While the core of this exoplanet population still lies at separations of a few AU, comparable to the lower-mass Jovian planets, a strong over-density of closer-in planets and brown dwarfs (a$_\mathrm{pl}\sim0.01$--0.1~AU) is seen among the sample of multiple-star systems (magenta), corresponding to the population of massive, small-separation sub-stellar companions in binaries highlighted previously from Figure~\ref{f:planet_mass-sma}. The minor peak at even tighter separations around single hosts corresponds to the sample of extremely short-period brown dwarfs found around white dwarfs discussed previously. At larger orbital distances, the directly-imaged population is subdued in the binary-star sample relative to closer-in planets and brown dwarfs, due to the effect explained above for systems with wide sub-stellar companions. 

Regarding the distribution of planet masses (Figure~\ref{f:kde_planet_prop}B), stellar binarity again seems to have no significant effect on the resulting masses for giant planets and brown dwarfs with separations larger than 0.5~AU (middle and bottom panels). At small semi-major axes (top panel), two sub-populations are observed, composed of the sub-Jovian planets with masses below 0.1~M$_\mathrm{Jup}$ forming the primary peak in the mass distribution, and a broader secondary population of giant planets and brown dwarfs. Again, we observe a relative over-abundance of binaries among the more massive planetary population on small semi-major axes, consistent with the findings deduced from our analysis as a function of planet orbital separation, and with the values reported in Table~\ref{t:results}.

\subsection{Planet Properties as a Function of Binary Properties}
\label{planet_prop_from_bin}

Based on our results from Section~\ref{bin_from_planet_prop}, suggesting that stellar multiplicity impacts the existence or properties of Jovian giant planets and brown dwarfs (M$_\mathrm{pl}>0.1$~M$_\mathrm{Jup}$) on semi-major axes within 0.5~AU, we further investigate the properties of these sub-stellar companions as a function of binary properties and the statistical significance of these results. We will not look in more details at other planetary systems as the previous analyses revealed no significant effect of binarity on these planetary populations. 

We assess the effect of binary separation by comparing the distribution of properties for close-in giant planets and brown dwarfs in binaries as a function of the orbital distance to outer stellar companions. Based on the size of this subset (66 sub-stellar companions), we arbitrarily define ranges of $<$250~AU, 250--1000~AU and $>$1000~AU in binary separation $\rho_\mathrm{bin}$, dividing this sample into roughly evenly-populated bins with 22, 24 and 20 systems, respectively. For hierarchical triple systems in which the planetary host star is in an inner tight binary, we only consider the close binary companion, as the outer tertiary component is unlikely to have a significant effect on the planetary system compared to the nearby stellar component. For triple systems with a planet host star widely separated from a closer binary, we count this outer binary as a single companion, using the mean separation between the planet host and the distant sub-system. Individual binary systems may be counted more than once, however, if several sub-stellar companions with masses larger than 0.1~M$_\mathrm{Jup}$ around found within 0.5~AU around the same star.

\begin{figure}
\begin{center}
\includegraphics[width=\textwidth]{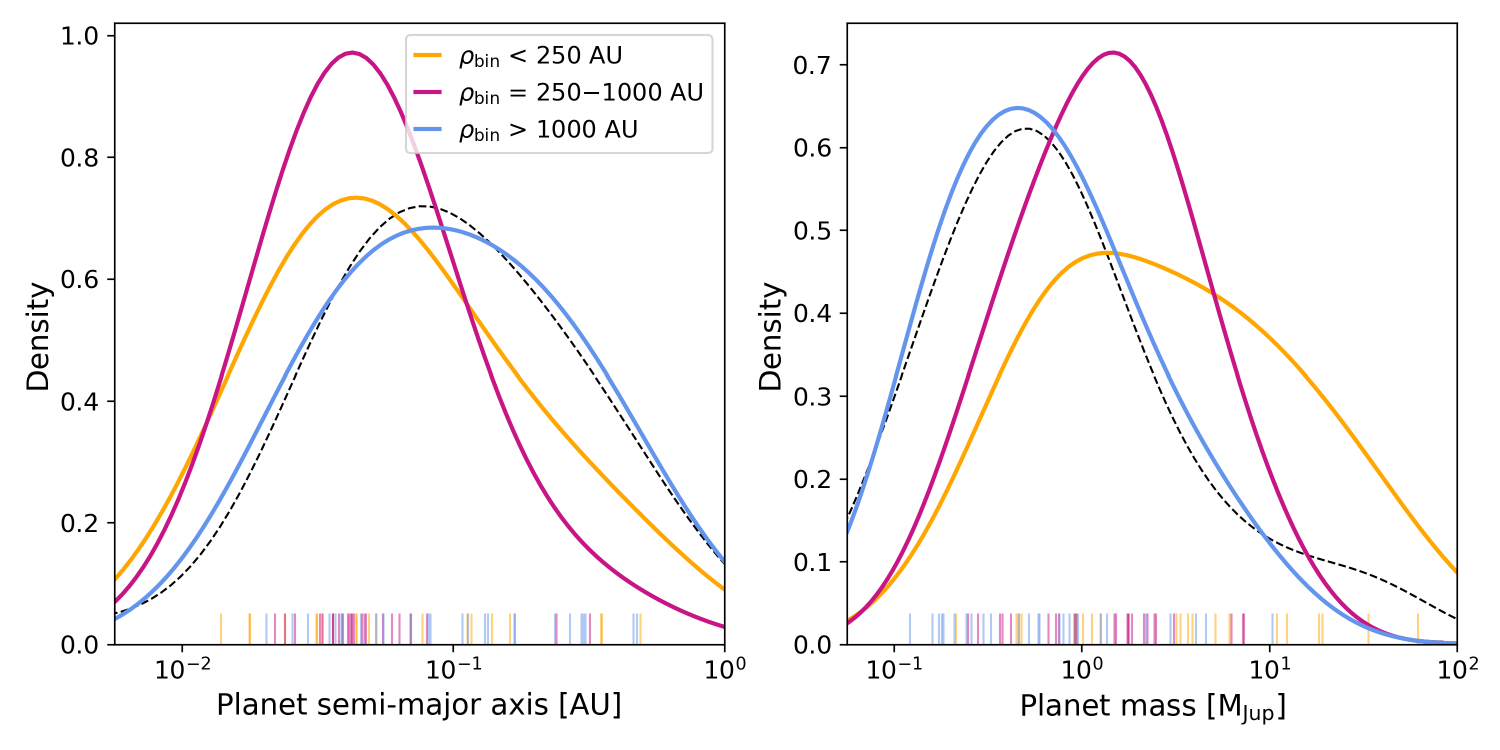}
\end{center}
\vspace{-0.3cm}
\hspace{0.25\textwidth}\textbf{(A)\hfill(B)}\hspace{0.25\textwidth}
\caption{Planet properties as a function of binary separation ($\rho_\mathrm{bin}$) for all planets with masses above 0.1~M$_\mathrm{Jup}$ and semi-major axes within 0.5~AU, corresponding to the binary systems plotted in the magenta distribution in Figure~\ref{f:bin_sep_distr}. KDEs of planetary semi-major axis are shown in panel A, and distributions of planet masses are shown in panel B. The dashed black lines show the distributions for planets in the mass and semi-major axis ranges found to be orbiting single stars.}
\label{f:kde_bin_sep}
\end{figure}

Figure~\ref{f:kde_bin_sep} shows KDEs of the planet semi-major axes (Figure~\ref{f:kde_bin_sep}A) and mass (Figure~\ref{f:kde_bin_sep}B), comparing planets and brown dwarfs in the short (yellow), intermediate (magenta) and wide (blue) binary separation ranges to those around single stars (dashed black line). Despite the small sample size available for this restricted planetary population, clear trends are visible in these figures. In particular, the subset of sub-stellar companions in extremely widely-separated binaries ($\rho_\mathrm{bin}>1000$~AU) shows very similar distributions in planetary semi-major axis and mass to planets and brown dwarfs found in single-star systems. In contrast, sub-stellar companions found in tighter stellar binary systems appear to have smaller semi-major axes and higher masses. The previously-noted overabundance of massive, close-in giant planets and brown dwarfs in binaries is hence primarily found in $<$1000~AU binary systems. We highlight, in particular, that from the 9 massive (M$_\mathrm{pl}>7$~M$_\mathrm{Jup}$), close-in (a$_\mathrm{pl}<0.5$~AU) giant planets and brown dwarfs found in binaries, 8 are in binaries with separations $<$1000~AU, from which 6 have binary separations $<$250~AU. While these rare sub-stellar companions only represent $\sim$2$\,\%$ of the full exoplanet sample, these systems make up about $10\,\%$ of the 64 binaries with separations under 250~AU identified for the full catalog of planet hosts. We further assess the significance of these results by performing two-sided Kolmogorov-Smirnov tests comparing each sub-population of planets in binaries to the sample of planets around single stars (dashed black lines). We are thus testing the null hypothesis that the samples are drawn from the same distribution, and use a threshold of 0.05 on the resulting p-values. We found that the null hypothesis could be rejected for the distributions of planet masses and semi-major axes in short and intermediate-separation binaries ($\rho_\mathrm{bin}<1000$~AU; yellow and magenta curves), but not for sub-stellar companions in very wide binaries (blue), confirming that the above findings are statistically significant (p-values of 0.027 and 0.0003 for the planet semi-major axes in short and intermediate-separation binaries, respectively, compared to 0.842 for wider binaries; p-values of 0.005 and 0.013 for the planet masses in short and intermediate-separation binaries, and 0.470 for wide binaries). This result further suggests that close and intermediate-separation ($<1000$~AU) binary companions have strong effects on the final semi-major axes of massive planets and brown dwarfs, whereas planetary systems in very wide ($>1000$~AU) binaries are more likely to evolve as independent stars.

\begin{figure}
\begin{center}
\includegraphics[width=\textwidth]{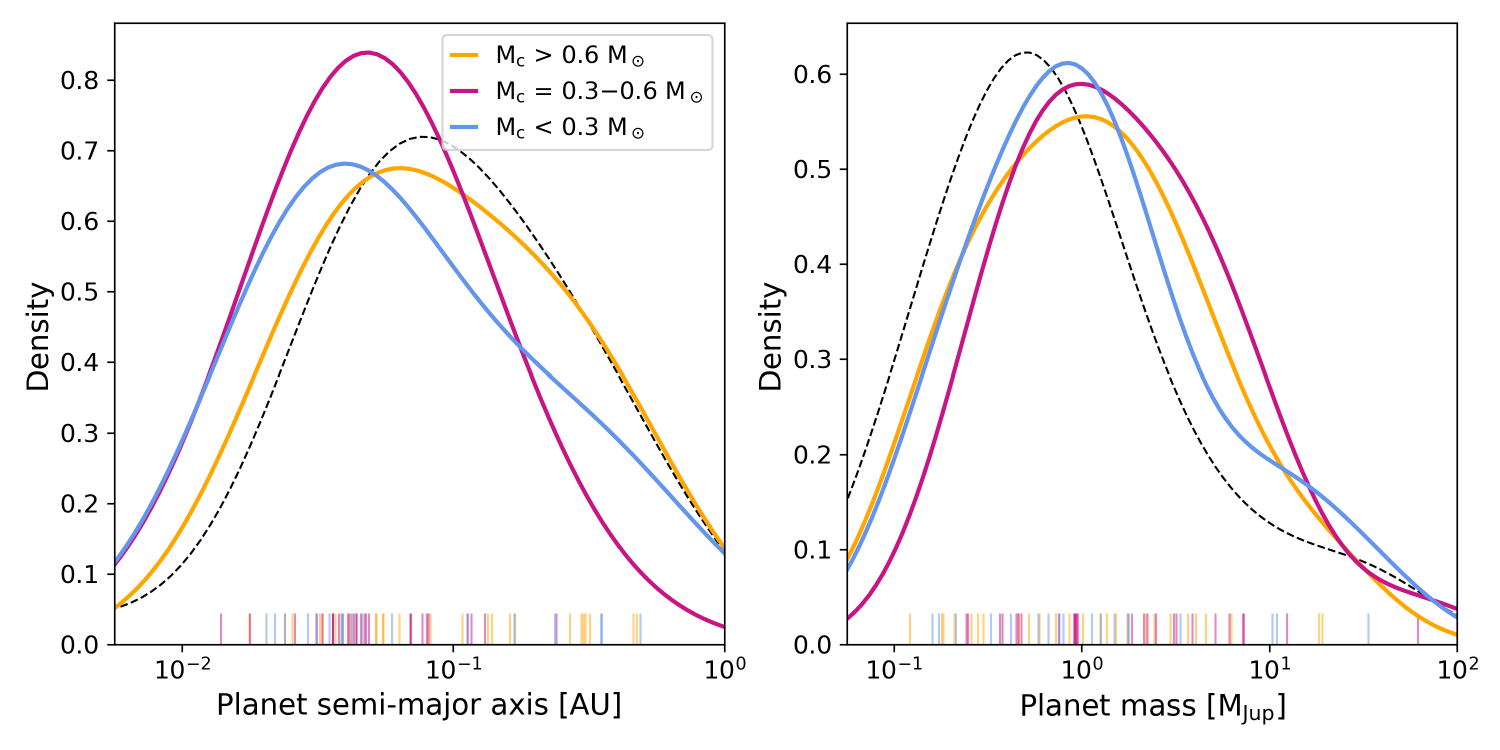}
\end{center}
\vspace{-0.3cm}
\hspace{0.25\textwidth}\textbf{(A)\hfill(B)}\hspace{0.25\textwidth}
\caption{Same as Figure~\ref{f:kde_bin_sep} dividing the sample of binary stars by companion mass (M$_\mathrm{c}$). For triple systems with a tight binary on a wide separation from the planet host, the total mass of the outer sub-system is considered. The dashed black lines show the distributions for planets in the mass and semi-major axis ranges found to be orbiting single stars.}
\label{f:kde_bin_mass}
\end{figure}

We also investigate potential trends of planet and brown dwarf properties as a function of binary companion mass, M$_\mathrm{c}$. As for the binary separation, we divide the available sample into bins of $<$0.3~M$_\odot$, 0.3--0.6~M$_\odot$ and $>$0.6~M$_\odot$. Triple systems are treated similarly as in the previous analysis, using the total mass of the outer components in the case of tight binaries on wider separations from the planet hosts. Figure~\ref{f:kde_bin_mass} shows the resulting distributions of planet semi-major axis (Figure~\ref{f:kde_bin_mass}A) and mass (Figure~\ref{f:kde_bin_mass}B) for the various stellar companion mass bins, together with the overall distributions of single-star planetary systems (dashed black line). Unlike Figure~\ref{f:kde_bin_sep}, no clear trend is observed with binary companion mass. The only marginal tendency is a rather comparable distribution between the planet orbital distances of single-star systems and the binaries with the most massive companions (yellow). Kolmogorov-Smirnov tests performed on these sub-samples confirmed that the planet separation distribution was statistically different from the single-star planetary population for binary systems with companion masses below 0.6~M$_\odot$ (p-values of 0.001 and 0.021 for binary companions in the intermediate and low mass bins, respectively; p-value of 0.859 for high-mass binary companions). However, this effect is mostly due to the fact that most stellar companions in this bin are in fact very distant, two-component companions from triple systems, thus increasing the adopted companion mass, and correspond for the major part to the systems with separations $>$1000~AU that were found to match the single-planet population. Kolmogorov-Smirnov tests could not reject the null hypothesis when comparing the masses of planets and brown dwarfs in various types of binaries to single-star systems, nor was any evidence found that sub-stellar companions in binaries with various stellar companion masses come from different populations (p-values $>$ 0.15 in all cases). Overall, the excess of smaller-separation and higher-mass giant planets and brown dwarfs in binaries appears to be distributed across the different binary mass bins defined, with no robust trend with stellar companion mass.

\section{Discussion}
\label{discussion}

In this work, we performed analyses of planetary populations as a function of multiplicity over all spectral types for hosts to exoplanets and brown dwarf companions. This section similarly presents discussions of our results across all types of stars, without distinguishing between massive stars, Sun-like stars and M dwarfs, or main sequence and evolved stars (sub-giants, giants or white dwarfs), unless explicitly stated otherwise. We note however that only 28 of our stellar hosts (out of 938, i.e. $<3\,\%$) are massive BA stars or white dwarfs, from which only 5 A stars were found to be in multiple systems (i.e. $\sim$2$\,\%$ of the binary sample). Excluding these systems would thus make little difference in the observed results and trends. While giants and sub-giants represent a more consequent fraction of the sample of host stars ($\sim$25$\,\%$ of the FGK hosts), a sizable number of our host stars have no luminosity class (giant/sub-giant vs. main sequence) in the spectral types gathered from the considered exoplanet catalogs or Simbad (e.g. numerous Kepler/TESS/WASP targets). We are therefore not able to strictly discuss main sequence stars separately, and our conclusions include a range of stellar masses and a mixture of stellar evolutionary stages.

\subsection{Stellar Mass Function and Multiplicity}

Figure~\ref{f:SpT_histogram} shows the distribution of spectral types from our planet host sample, divided between those identified in visual binaries or multiples (magenta) and seemingly single stars (blue), and compared to the identified stellar companions (yellow). Absolute numbers are provided at the top of each bar. In addition, each color-coded histogram is normalized so that the sum of the bars in a given color add up to 1, i.e. the height of each bar on the y-axis gives the relative contribution from that spectral type towards to full considered sub-sample.

Comparing the single and binary stars from our planet hosts, the subset of binary hosts contains a larger relative fraction of massive A, F and G stars, with a smaller contribution from lower-mass K and M dwarfs, as demonstrated by the turnover in the relative heights of the magenta and blue bars from G to K spectral types.
This is consistent with the well-known trend of decreasing binary rate with decreasing stellar mass, dropping from $\sim$70$\,\%$ for B and A stars \citep{Kouwenhoven2007} to around $50\,\%$ for Sun-like stars \citep{Raghavan2010}, and about 30$\,\%$ for M dwarfs \citep{Janson2012,Winters2019,Ward-Duong2015}.
While our survey results were not corrected for incompleteness and additional binaries may be missing from our compilation (see Section~\ref{completeness}), our ability to retrieve wide stellar companions for our sample can be assumed to be rather independent of the host spectral types. 
With raw binary fractions of $37.8\pm6.5\,\%$, $24.3\pm2.6\,\%$, $22.4\pm2.7\,\%$ and $15.7\pm3.1\,\%$ for F, G, K and M stars, respectively, these results thus suggest that the population of planet-bearing stars is representative of the relative multiplicity output of stellar formation across the stellar spectral sequence. However, without robust completeness corrections, we are not able to determine whether the differences between our observed raw fractions and overall stellar multiplicity rates are due to missing binaries in our samples or to the fact that stars hosting planets and brown dwarfs are truly less commonly found in binary-star systems.

The distribution of spectral types from companions, on the other hand, peaks strongly towards low-mass M dwarfs (yellow), which represent over 65$\,\%$ of our sample of stellar companions. In fact, this resembles closely the stellar initial mass function, with M dwarfs being the most abundant types of stars \citep{Chabrier2003,Bochanski2010}. This indicates that planet hosts in multiple systems are more often the most massive component of stellar binaries. The feature is partly due to a selection effect, as lower-mass stars are often too faint to be included in target samples for exoplanet campaigns \citep{Eggenberger2010}.
Nonetheless, although Earth to Neptune-sized planets are more abundant around M dwarfs \citep{Mulders2015}, giant planet formation is thought to be more efficient around more massive stars \citep{Mordasini2018}, and giant planets are indeed observed to be more frequent around higher-mass stars \citep{Bonfils2013,Vigan2020}. Given that binary systems seem to preferentially host giant planets based on our results, it is not surprising that most planet hosts in multiple systems would be the most massive stellar component in these hierarchical systems.

\subsection{Completeness and Survey Limitations}
\label{completeness}

As mentioned previously, the (in)completeness of our multiplicity search was not accounted for in the results presented in Section~\ref{results}, as corrections of observational biases are beyond the scope of this work. We may nonetheless look at the properties of our detected systems to understand what biases might lie in our gathered sample.

\begin{figure}
\begin{center}
\includegraphics[width=0.62\textwidth]{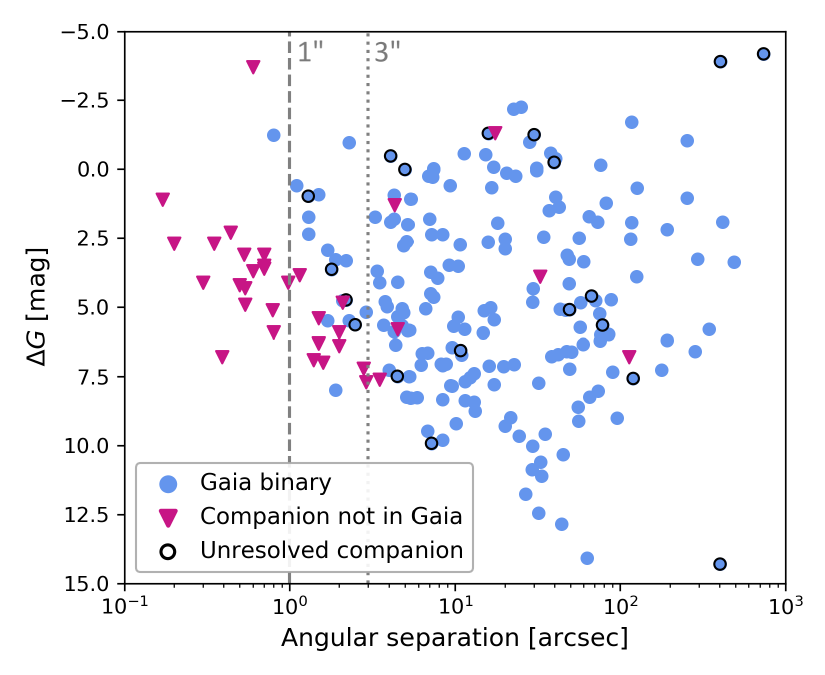}
\end{center}
\caption{Completeness of the \textit{Gaia}~DR2 binary search showing \textit{G}-band magnitude differences against angular separations for all companions retrieved in \textit{Gaia} (blue circles). Detected companions known to be themselves close binaries unresolved in\textit{Gaia} are marked with black circles.
Known companions not recovered in the \textit{Gaia}~DR2 catalog are shown in the magenta triangles, with plotted magnitude differences corresponding the lower limits in the \textit{Gaia} \textit{G}-band. The dashed and dotted gray lines show inner working angles of 1~arcsec and 3~arcsec, respectively.}
\label{f:gaia_completeness}
\end{figure}

Figure~\ref{f:gaia_completeness} shows the angular separation and \textit{Gaia} \textit{G}-band magnitude difference for every visual companion, relative to the planet host star it is bound to. Blue circles represent companions successfully retrieved in \textit{Gaia}~DR2. Binary components which are themselves known to be unresolved binaries are marked with black rings. Magenta triangles correspond to companions known from the literature but undetected in \textit{Gaia}. As a $\Delta G$ magnitude difference is unavailable for these systems, the plotted magnitudes correspond to contrasts in various visual or infrared filters, and thus correspond to lower limits compared to the expected magnitude difference values in the \textit{G}-band. Our observed recoverability for binaries is consistent with the estimated \textit{Gaia} completeness to close binaries \citep{Ziegler2018}: near equal-brightness binaries ($\Delta G < 2$~mag) are consistently retrieved from separations of 1~arcsec (dashed line), binaries down to around $\Delta G = 6$~mag are typically recovered at separations of $\sim$3~arcsec (dotted line), and wider systems are subject to \textit{Gaia}~DR2 completeness down to the limiting magnitude of $G\sim21$~mag of the \textit{Gaia}~DR2 survey.

\begin{figure}
\begin{center}
\includegraphics[width=0.62\textwidth]{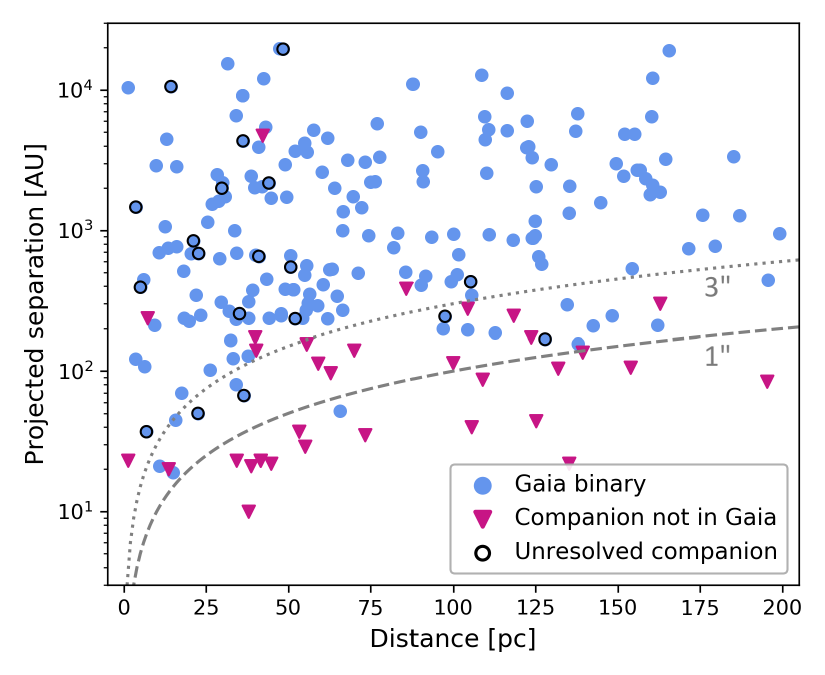}
\end{center}
\caption{Physical projected separations against distance all identified stellar companions, with the same symbols and color-codes as in Figure~\ref{f:gaia_completeness}, highlighting the completeness of our \textit{Gaia} binary search as a function of distance to the Sun. The dashed and dotted gray lines show inner working angles of 1~arcsec and 3~arcsec, respectively.}
\label{f:gaia_completeness2}
\end{figure}

A significant number of known tight binaries with angular separation $<$1~arcsec (magenta triangles) were not recovered in \textit{Gaia}, only known thanks to high angular-resolution imaging campaigns. As only a small fraction of our hosts stars have been targeted by such dedicated imaging programs, these results indicate that additional unresolved sub-arcsecond systems may still be hidden among our exoplanet host sample. In particular, the 27 binaries for our sample with projected separations $<$100~AU have a median angular separation of 0.7~arcsec, and such systems are therefore for the most part not recoverable in \textit{Gaia}. Studies like \citet{Kraus2016} or \citet{Furlan2017} have identified numerous optical candidate companions to Kepler hosts stars on small angular separations, but additional observational epochs are required to confirm or refute the bound nature of most of these candidates.

A shortfall of close binaries ($<$50--100~AU) among planet hosts has been vastly reported in observational surveys \citep{Bergfors2013,BonavitaDesidera2020,Kraus2016,MoeKratter2019,Roell2012,Wang2014}. This feature is generally attributed to a hindrance of planet formation in very tight binaries, and is also predicted in theoretical models \citep{ThebaultHaghighipour2015}.
However, our survey is highly incomplete out to separations of hundreds of AU and thus cannot be used to probe this feature. 
Indeed, the resolving limit of $\sim$1--3~arcsec in our \textit{Gaia} search corresponds to projected separations of 200--600~AU for the most distant stars in our study (200~pc). This effect is illustrated in of Figure~\ref{f:gaia_completeness2}, which plots the physical projected separation of all identified binaries as a function of distance from the Sun. Detection limits corresponding to inner working angles of 1~arcsec and 3~arcsec are marked with dashed and dotted lines, respectively. The figure clearly demonstrates that the range of probed binary separation is strongly affected by the distance to each star.
Our compilation is only sensitive in \textit{Gaia} to binary separations below 100~AU for targets out to 30~pc ($\sim20\,\%$ of the sample), and only data from heterogeneous high-angular resolution programs have allowed the detection of such systems beyond 100~pc.

\subsection{Impacts of Multiplicity on Exoplanets}

\subsubsection{No Influence on Low-Mass Planets}

We found that small planets with masses below 0.1~M$_\mathrm{Jup}$ have a significantly lower raw binary rate ($16.6\pm1.7\,\%$) than more massive Jovian planets ($25.5\pm1.8\,\%$ for planets above 0.1~M$_\mathrm{Jup}$ throughout the brown dwarf mass range), an offset with a 3.6-$\sigma$ significance. While these numbers certainly suffer from inherent and observational biases as discussed in Section~\ref{completeness}, it is reasonable to assume that these biases do not affect hosts to different types of planets differently. Indeed, the transit and radial velocity surveys that yield the detection of these planets are partially subject to the same inherent selection biases as campaigns discovering more massive planets with the same methods. As a result, we consider that the observed trend of lower multiplicity fraction for sub-Jovian planets is a real feature.
Furthermore, terrestrial and Neptunian planets are often found in tightly-packed multiple-planet systems \citep{Mayor2011}. The fact that such planets are less frequently seen in hierarchical star systems is thus also consistent with the observation that multi-planet systems are less commonly found in stellar binaries. 

These results, however, may be a direct consequence of the lower binary frequency of M dwarfs compared to more massive stars. Since low-mass M stars host $\sim$2--3 times more close, small planets than Sun-like stars \citep{Howard2012,Mulders2015}, and rarely harbor giant planets \citep{Bonfils2013}, the majority of small sub-Jovian planets and high-order multi-planet systems are therefore found around M dwarfs. The intrinsic lower stellar multiplicity rate of M dwarf could hence be responsible, at least partly, for the observed trends. Nonetheless, \citet{MoeKratter2019} found that the biases from stellar companions against the detection of planets are higher for F and G stars than M dwarfs. This trend is rooted on the suppression of planet formation in close binaries and bright stellar companions preventing transit detections. This suggests that the observed differences in raw binary fractions between sub-Jovian and giant planet systems would likely be increased after accounting for these biases, and we conclude that low-mass planets and tightly-packed systems with multiple small planets are truly less commonly found in hierarchical stellar systems.

\subsubsection{The Excess of Massive Close-in Planets and Brown Dwarfs in Binaries}
\label{massive_planets}

The substantial prevalence of short-orbit massive planets and brown dwarfs around members of binary stars was first noted by \citet{ZuckerMazeh2002}, and later confirmed by numerous observational studies \citep{Eggenberger2004,DesideraBarbieri2007,Mugrauer2007}. More recently, the Friends of Hot Jupiters survey reported an enhancement of binary frequency for stars hosting hot Jupiters, with a binary rate 3 times higher that for field stars over the separation range 50--2000~AU \citep{Ngo2016}. \citet{Fontanive2019} established the continuity of this trend to the most massive giant planets and brown dwarfs ($>$7~M$_\mathrm{Jup}$) found within $\sim$1~AU, constraining the binary frequency of such systems to be around $80\,\%$ between 20 and 10\,000 AU, a result further validated statistically in \citet{MoeKratter2019}. Results from these studies demonstrate that stellar companions play an important role in the formation and/or evolution of these rare planetary systems. These findings also suggest that the influence of binary companions is strengthened for higher-mass close-in exoplanets and sub-stellar companions, and that this effect may be magnified for sub-stellar companions on even tighter orbits. 

While the work presented here did not allow us to place any such frequency constraints, the intrinsic tendencies with planet mass and separation observed in previous studies are confirmed in our compilation. Indeed, we observed a larger relative fraction of Jovian planets in binaries within 0.5~AU than for the bulk of the Jovian planet population around the snow line ($\sim$1--5~AU). This relative frequency was found to further increase when focusing exclusively on the most massive planets and brown dwarfs.
These trends suggest that stellar multiplicity affects the orbital separation of massive giant planets. The presence of an outer wide companion would hence allow for the inner sub-stellar companion to reach closer-in semi-major axes than planets of similar masses orbiting single stars, onto an orbital separation regime where essentially no planets around single stars are observed \citep{Fontanive2019}. 

The influence from outer stellar companions shows a possible dependence on binary separations. Stellar companions on separations of the order of thousands of AU seem to have no significant effect on the demographics of planetary systems, with similar distributions observed between the masses and semi-major axes of planets and brown dwarfs in such binaries and around single stars. In contrast, the most massive, close-in giant planets and brown dwarfs in binaries, in the most extreme planetary configurations, are all in rather tight binaries, with separations of tens to a few hundreds of AU compared to a mean $\sim$600~AU for the full binary sample (likely a direct consequence of uncorrected incompleteness biases as discussed in Sections~\ref{completeness} and~\ref{binary_sep}). This is consistent with the observed peak in binary separation from \citet{Fontanive2019} for such systems ($\sim$250~AU), and further supports the idea that additional binaries may remain undiscovered in our probed sample on this separation range. 

On the other hand, no robust dependence of binary influence on stellar companion mass was seen in our results. This is not surprising since the gravitational pull from a companion scales with M$_\mathrm{c}/\rho_\mathrm{bin}^2$ (where M$_\mathrm{c}$ is the companion mass and $\rho_\mathrm{bin}$ the binary separation). As the companion masses span a range of about one order of magnitude, compared to over three orders of magnitude for the separation (which is then squared), binary separation is thus expected from physical arguments to have larger impact on the circumstellar planetary system.

\subsubsection{Very Close Binaries in Triple Systems}

The vast majority of main sequence stars in spectroscopic binaries are known to be the inner binaries of hierarchical triple systems. \citet{Tokovinin2006} demonstrated that 96\% of binaries with orbital periods below $\sim$3~days have tertiary stellar companions. The occurrence of outer components for these systems is found to steadily decrease with inner binary period, falling to a rate of 34\% triple systems for spectroscopic binaries with periods of 12--30~days. The excess of tertiary companions has been argued \citep{FabryckyTremaine2007,NaozFabrycky2014} to allow for the migration of the inner companions via Kozai-Lidov oscillations in misaligned triples \citep{Kozai1962,Lidov1962}. Alternatively, these close binary companions have been suggested to form via disk fragmentation and migration within the circumstellar disk of the primary star \citep{MoeKratter2018}. The substantial mass required to form and drive inward such massive inner companions can simultaneously form additional tertiary companions, leading to such systems being often in triple-star configurations. These outer components could then allow for more extreme migrations of the inner companions, leading to the observed negative correlation between inner binary period and triple architecture frequency \citep{MoeKratter2019}.

\citet{Fontanive2019} studied hosts to close giant planets and brown dwarf companions with masses of 7--60~M$_\mathrm{Jup}$, inferring a tertiary companion fraction comparatively high to the spectroscopic binaries from \citet{Tokovinin2006}. This population of sub-stellar companions corresponds to the most massive, short-separation systems found to be predominantly in hierarchical stellar structures in this work. \citet{MoeKratter2019} further confirmed this excess of triple occurrence rate to be a statistical, real feature, as well as to be measurably higher than for genuine hot Jupiters ($<$4~M$_\mathrm{Jup}$) surveyed by \citet{Ngo2016}. The similar demographics between these brown dwarf desert systems and stellar spectroscopic binaries argues for a common origin for the inner companions from \citet{Tokovinin2006} and \citet{Fontanive2019}, indicating that these inner giant planet and brown dwarf companions extent the population of triple stellar systems to sub-stellar masses for the secondary components of the inner binaries.

\begin{figure}
\begin{center}
\includegraphics[width=0.6\textwidth]{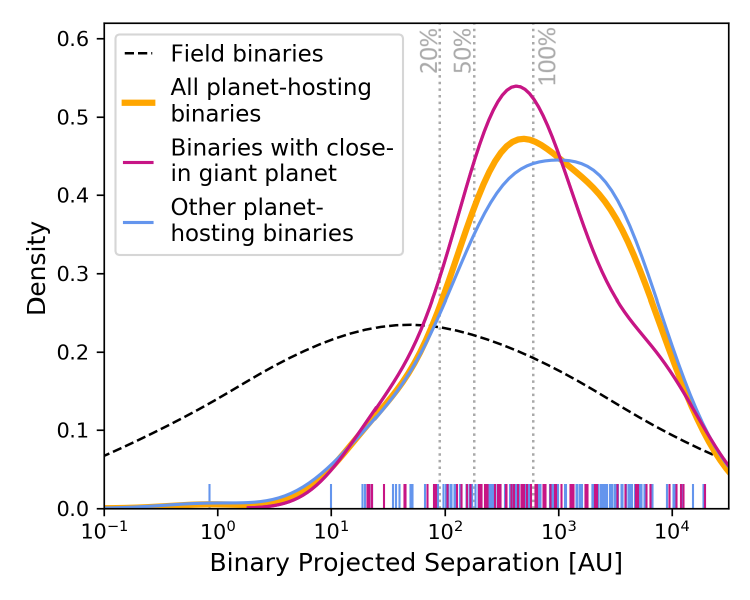}
\end{center}
\caption{Binary separation distribution comparing the full sample of planet-hosting binaries (thick blue line) to stellar binaries in the Solar-neighborhood (dashed black line) from \citet{Raghavan2010}. The sample of planet-bearing multiples is further divided between systems hosting a giant planet of mass $>$0.1~M$_\mathrm{Jup}$ within 0.5~AU (magenta) and all other systems (yellow). Multi-planet systems with planets falling into the two planetary categories are counted towards the close-in giant planet subset. Planet host stars in close binaries with an outer tertiary companion are plotted as the inner binary only. Triple systems composed of a planet host and a wide tighter binary are counted as a binary system using the mean separation to the distant sub-system. The dashed vertical gray lines show the projected separations probed for the closest 20$\,\%$, 50$\,\%$ and 100$\,\%$ of our sample for angular separations of 3~arcsec.}
\label{f:bin_sep_distr}
\end{figure}

\subsubsection{The Effect of Binary Separation}
\label{binary_sep}

In Figure~\ref{f:bin_sep_distr}, we compare the distributions in projected separation of the planet-hosting wide binaries (solid yellow line) gathered in this work, and the Solar-type field binaries (dashed black line) from \citet{Raghavan2010}. For multi-planet systems with planets or brown dwarfs falling into the two planetary categories considered (55 Cancri, HD 38529 A, Upsilon Andromedae A, WASP-8), we count the binaries only once, towards the close-in giant planet subset. Triple systems are accounted for in the same way as in Section~\ref{planet_prop_from_bin}.

The binary separations of the planet-bearing systems appears to peak at significantly larger values, with a peak around 600~AU compared to $\sim$50~AU for field binaries. Field binaries also show a much broader distribution, with a log-width of 1.70 compared to 0.75 for planet-hosting multiples. A Kolmogorov-Smirnov test confirms that the raw observed distributions are indeed statistically different, with a p-value for the null hypothesis that they are drawn from the same distribution $<10^{-5}$.
These differences are primarily due to the incompleteness of our compilation on short binary separations as discussed in Section~\ref{completeness}. Furthermore, the field binaries from \citet{Raghavan2010} include unresolved companions detected by spectroscopic techniques or proper motion accelerations. Such systems are not detectable with visual detection methods and a number of such tighter binaries could remain undetected in our studied sample. 

The dashed lines in Figure~\ref{f:bin_sep_distr} show the projected separations probed for the closest 20$\,\%$, 50$\,\%$ and 100$\,\%$ of our sample with angular separations of 3~arcsec, our adopted completeness limit for \textit{Gaia}. Our observed peak in the separation distribution (600~pc) roughly coincides with our inner completeness limit for the full sample (see Section~\ref{completeness}). This strongly suggests that a number of undiscovered binaries with separations of tens to hundreds of AU may still lie in our sample. For example, the planet host stars DMPP-3~A and HD~59686 from our exoplanet compilation were both found through significant radial velocity trends to have close stellar companions at 1.22~AU \citep{Barnes2020} and 13.56~AU \citep{Ortiz2016}, respectively. These companions have never been resolved to this date, and these systems were thus counted as single in the context of this work, which only considered visual, astrometrically-confirmed systems. Current high-angular resolution efforts and complementary detection methods, probing smaller binary separation ranges, must thus be pursued to obtain a more complete picture of the multiplicity of exoplanet host stars, and understand the true effect of tight binary companions on the formation and evolution of extra-solar planets and brown dwarf companions. 

Figure~\ref{f:bin_sep_distr} also shows the distribution of binary separation, dividing the sample between binaries hosting a giant planet or brown dwarf within 0.5~AU (magenta) and the all other systems (blue).
The sample of binaries hosting a short-period gas giant appears to be on somewhat smaller binary separations than the remaining planet-hosting multiples, with logarithmic means shifting from $\sim$500~AU to $\sim$700~AU between the two samples, and a slightly tighter distribution for the former subset. This is representative of our results from Section~\ref{planet_prop_from_bin}, which showed an enhanced relative fraction of shorter-separation binaries for systems with close-in planets, an effect that is further magnified for the most massive planets and brown dwarfs on very tight orbits. These results are in agreement with previous observations \citep{DesideraBarbieri2007} that found massive planets in short period orbits to be in most cases around the components of rather tight binaries.
Finally, the larger relative number of very wide binaries ($>$1000~AU) for hosts to lower-mass and larger-separation planets is also consistent with the rest of our results. We indeed found that such widely-separated binaries generally do not impact the planet properties (see Figure~\ref{f:kde_bin_sep}), and observed small and wide-orbit planets to not be significantly affected by the presence of stellar companions, compatible with the idea that most such planets are only found in very wide binaries or around single stars. \citet{DesideraBarbieri2007} similarly concluded that the properties of exoplanets orbiting components of very wide binaries are compatible with those of planets orbiting single stars.

\subsection{Implications for Formation Mechanisms}

The final architectures of planetary systems around members of binary stars strongly depend on how the presence of a close massive body impacts standard formation and migration processes, through its efficiency to alter the local disk environment, accretion rates, or tidal interactions between planets and the host star. The population trends highlighted throughout this study might provide new clues and insights into the effects of stellar companions on planet and brown dwarf formation and evolution. 

Our findings show that binarity has little effect on the distributions of planet mass and semi-major axis for the population of sub-Jovian planets found inside $\sim$1 AU. The impact of stellar duplicity on short binary separations ($<$50~AU) remains to be fully understood theoretically and better constrained observationally, as such very tight binaries would be more amenable to influence disks inner regions crucial for small planet formation and stable orbital behaviors. Nonetheless, our results demonstrate that sub-Jovian planets that form in binaries with separations of hundreds of AU are consistent with the population of single-star planets. This suggests that stellar multiplicity does not need to be extensively accounted for in order to reproduce the core of this population of small planets, either completely inhibiting the formation or survival of such planets, or having no visible effect on the demographics of successfully formed planets.

In contrast, we found that the populations of intermediate-mass giant planets (M$_\mathrm{pl}=0.1$--7~M$_\mathrm{Jup}$) and high-mass sub-stellar companions (M$_\mathrm{pl}>7$~M$_\mathrm{Jup}$) show different statistical properties between single-star systems and hosts with stellar companions. Small-separation planets and brown dwarf companions within the snow line ($<$1--3~AU) were found to have somewhat larger masses and/or tighter separations when in binary stars. This trend is enhanced for the most massive and closest-separation sub-stellar companions, that also have inflated raw stellar multiplicity rates compared to lower-mass and wider planets, consistent with previous studies \citep{Fontanive2019,MoeKratter2019}. This strongly indicates that the identified stellar binary companions likely affect the formation and/or migration of these massive sub-stellar objects, either allowing for more massive planets to exist at similar separations than planets around single stars, or enabling similar-mass planets to form or migrate to shorter semi-major axes in stellar binaries.

Understanding the true nature and extend of these effects is a challenging task. Unfortunately, available data provide little insight at this stage into the details of the possible underlying processes, with few prospects to disentangle between planetary formation and evolution, and missing information about most orbital elements for binary orbits. Likewise, modeling the formation and evolution of planets in binaries requires to explore a very wide parameter space, including binary separations, mass ratios, inclinations and eccentricities. The large variety of possible binary configurations likely impacts the existence and properties of planetary systems differently for each combination of these key binary parameters, from detrimental to perturbing or even favorable effects.

For circumstellar planets, which represent the focus of this study, a nearby stellar companion is expected to primarily affect the outer parts of typical planet formation locations, where the gravitational influence from the stellar companions will be enhanced. The outskirts of protoplanetary disks are believed to predominantly harbor more massive planets, with mostly rare, cold Jovian planets predicted beyond a few tens of AU in the core accretion paradigm \citep{Emsenhuber2020}, and the formation of massive planets and brown dwarfs by gravitational disk fragmentation occurring preferentially in the cool outer regions of disks, from separations of several tens to hundreds of AU \citep{Rafikov2005,Hall2017}. Following this reasoning, giant planets and brown dwarfs forming at large orbital separations are thus more likely to be affected by the presence of an outer star in the system than small planets forming and accreting within a few AU from the host star. The observed population of wide-orbit ($>$10~AU) planets and brown dwarfs was found to have a lower raw binary rate than similar-mass sub-stellar companions on shorter orbits, and no significant differences were observed in planetary properties between single and multiple-star systems. The effect from outer companion stars would thus likely be in facilitating inward migration processes, bringing massive giant planets and brown dwarfs onto extremely tight orbits typically unreachable in single-star environments, via e.g. the Kozai-Lidov mechanism \citep{NaozFabrycky2014,Winn2010} or other triggered dynamical perturbations.

Alternately, binarity could impact separate planet formation channels differently, i.e., influencing the conditions for gravitational disk instability, but with little effect on the results of core accretion mechanisms if they proceed, thus affecting the very most massive planets and brown dwarfs only. For example, the presence of a nearby companion star within $\sim$100--300~AU could tidally truncate protoplanetary disks \citep{Kraus2012} and lead to faster disk dissipation rates \citep{Muller2012}. This effect would be particularly problematic for giant planet formation by core accretion, which requires significantly longer timescales to operate than disk fragmentation. Formation by core accretion would therefore only take place if the outer companion has little effect on the disk and forming planetary system, thus not  significantly impacting the final planet properties compared to single star conditions. Similarly, binary companions have been suggested to be able to trigger instabilities in otherwise stable disks \citep{Boss2006}, hence favorably modifying formation environments for \textit{in-situ} disk fragmentation, but inconsequential for core accretion. This idea is reconcilable with the high masses of the outlying population of sub-stellar companions observed to be predominantly in binaries, which seemingly formed differently from the population of lower-mass planets on similar orbits, most likely through gravitational disk fragmentation \citep{MoeKratter2019}.

Finally, our results regarding the separation distribution of binaries might help to narrow down the effect of at least one binary parameter. As mentioned previously, there is reliable observational evidence that close binarity ($<$50~AU) hinders planet formation around a host star \citep{Fontanive2019,Kraus2016,Wang2014}, although this feature could not be robustly investigated in the present work. We also found stellar multiplicity of very large separations (thousands of AU) to have no significant impact on observed planetary populations, suggesting that planet formation and evolutionary patterns in such systems behave similarly as around single stars. Intermediate separations, from several tens to a few hundreds of AU, therefore appear to be a key region of the parameter space to explore in order to further our perspective of exoplanets in stellar binaries.  Examinations of physical quantities in these systems such as binding energy may be especially interesting to study for a better physical understanding of the processes in play, and we particularly advocate for investigations to be conducted in the theoretical context of gravitational disk instability based on our results.

\vspace{2cm}
\section{Summary}
\label{conclusion}

In this work, we have compiled a sample of 938 stars hosting a total of 1316 extra-solar planets and brown dwarf companions, out to 200~pc. We searched for visual co-moving companions to these systems via an extensive search in the literature and using the \textit{Gaia}~DR2 catalog to identify common proper motion sources. This analysis yielded a total of 218 planet hosts in multiple-star systems, including 186 binaries and 32 hierarchical triple systems, with 10 newly-discovered binary companions and 2 new tertiary components. From these, 4 binaries and 1 triple system contain 2 planet-bearing stars. Stellar companions have masses ranging from the brown dwarf/star boundary at 0.07~M$_\odot$ up to 2.27~M$_\odot$, with separations ranging from $<$1~AU to 20\,000~AU with a median of $\sim$600~AU. 

Investigating our gathered sample of binaries, we found that:
\begin{enumerate}
    \item[$\bullet$] More massive planet hosts are more often part of multiple-star systems, consistent with the population of planet-bearing stars following the overall relative multiplicity outcome of stellar formation. 
    \item[$\bullet$] Planet hosts in multiple systems were also predominantly observed to be the most massive component of stellar binaries.
    \item[$\bullet$] A total of 27 binary systems have separations $<$100~AU, from which 20 have binary separations smaller than 50~AU, with 1 system in an extreme $<$1~AU configuration. Most of these close binaries, however, were only identified thanks to dedicated high-angular resolution campaigns, and could not, for the most part, be retrieved with the resolving limit of \textit{Gaia} (1--3~arcsec), in particular for the most distant targets in our sample. 
    \item[$\bullet$] As only a small fraction of planet hosts have been targeted by such imaging programs, a significant number of sub-arcsecond binaries and companions on separations of a few arcseconds could still be missing from our catalog, further supported by the concurrence of our measured peak in binary separation and our estimated \textit{Gaia} completeness limit.
\end{enumerate}

Assuming that the selection and observational biases lying in and limiting our gathered compilation of stellar binaries affect various subsets of planetary populations and planet hosts in a reasonably homogeneous way, we investigated possible correlations between planet properties and the existence and properties of outer stellar companions. Our main results are:
\begin{enumerate}
    \item[$\bullet$] From our identified sample of binary companions, we measured a raw multiplicity rate of $23.2\pm1.6\,\%$ for planet hosts. 
    \item[$\bullet$] Multi-planet systems were found to have a somewhat lower stellar duplicity frequency ($18.0\pm2.7\,\%$) compared to single-planet systems ($25.1\pm1.9\,\%$) with a 2.2-$\sigma$ significance. 
    \item[$\bullet$] Dividing the planet parameter space into various sub-populations, we found that giant planets and brown dwarfs with masses above 0.1~M$_\mathrm{Jup}$ have a substantially larger (3.6-$\sigma$) raw stellar multiplicity fraction ($25.5\pm1.8\,\%$) than lower-mass planets ($16.6\pm1.7\,\%$), consistent with the fact that these small sub-Jovian planets are typically organized in tightly-packed multi-planet systems.
    \item[$\bullet$] This trend appears to further increase up to about $\sim$30$\,\%$ for massive planet and brown dwarfs (M$_\mathrm{pl}>7$~M$_\mathrm{Jup}$) on very short orbital separations (a$_\mathrm{pl}<0.5$~AU), with the most massive and shortest-period sub-stellar companions almost exclusively observed in multiple-star systems. These results are consistent with previous studies of these populations \citep{Fontanive2019,MoeKratter2019}, which appear to follow the architectures of stellar spectroscopic binaries, systematically observed as part of hierarchical triple systems \citep{Tokovinin2006}. 
\end{enumerate}

In terms of planet properties, our results suggest that:
\begin{enumerate}
     \item[$\bullet$] Stellar duplicity has no significant effect on the demographics of low-mass planets (M$_\mathrm{pl}<0.1$~M$_\mathrm{Jup}$) or the core population of warm giant exoplanets on separations neighboring the snow line (a$_\mathrm{pl}>0.5$~AU).
     \item[$\bullet$] Only high-mass, small-separation planets were observed to have different distributions of planet properties between the subset of planets in binaries and single-star systems, with an over-density of planets and brown dwarfs of several Jupiter masses found on semi-major axes of $\sim$0.01--0.1~AU identified in multiple-star systems. \item[$\bullet$] These extreme planetary systems with few or no single-star analogues were predominantly found to be in rather tight binary configurations $<$1000~AU, and mostly on separations $<$250~AU for sub-stellar companions with masses $>$7~M$_\mathrm{Jup}$. These systems represent a sizable fraction of such tight binaries in our compilation ($\sim$10$\,\%$), despite the rarity of these planets and brown dwarfs in our overall exoplanet catalog ($\sim$2$\,\%$).
     \item[$\bullet$] In contrast, the subset of these planets in binaries with separations $>$1000~AU showed similar distributions in mass and semi-major axis to planets and brown dwarfs orbiting single stars. This indicates that short ($<$250~AU) or intermediate-separation ($<$1000~AU) binaries play a role the formation or evolution of these massive planets and brown dwarfs, but that very wide binaries do not influence the architectures of planetary systems.
      \item[$\bullet$] Binary companion mass, on the other hand, was found to have no significant effect on planetary properties.
\end{enumerate}

Between the upcoming generation of telescopes and future \textit{Gaia} data releases, the next decade promises to lead to unprecedented discoveries and new characterisation possibilities. These findings will arguably yield unparalleled information and new robust constraints on system architectures and population demographics, which will in turn provide key probes into formation histories and dynamical evolution processes. We hope that the gathered compilation of exoplanets in visual binaries will be useful to future studies in this constantly-growing research area, and will motivate the need to pursue existing campaigns searching for small-separation binary companions to known planetary systems. With a more comprehensive picture of stellar multiplicity on the separation ranges demonstrated here to remain highly incomplete, we will be able to confirm and better understand the tentative trends highlighted in this paper, and improve our fundamental understanding of stellar, sub-stellar and planetary formation and evolution.

\section*{Conflict of Interest Statement}

The authors declare that the research was conducted in the absence of any commercial or financial relationships that could be construed as a potential conflict of interest.

\section*{Author Contributions}

C.F. led this study, performing the catalog compilations and binary searches and leading the data analyses. D.B.G. contributed to the search for trends within the obtained catalogs, scientific interpretation of results, and writing of the paper.


\section*{Acknowledgments}
C.F. acknowledges support from the Center for Space and Habitability (CSH). This work has been carried out within the framework of the NCCR PlanetS supported by the Swiss National Science Foundation.
D.B.G. acknowledges support from the NASA ADAP award No. 80NSSC19K0532.
This research has made use of the NASA Exoplanet Archive, which is operated by the California Institute of Technology, under contract with the National Aeronautics and Space Administration under the Exoplanet Exploration Program.
This research has made use of the Exoplanet Orbit Database and the Exoplanet Data Explorer at exoplanets.org.
This work has made use of data from the European Space Agency (ESA) mission {\it Gaia} (\url{https://www.cosmos.esa.int/gaia}), processed by the {\it Gaia} Data Processing and Analysis Consortium (DPAC, \url{https://www.cosmos.esa.int/web/gaia/dpac/consortium}).
This research has made use of the SIMBAD database and the VizieR catalogue access tool, operated at CDS, Strasbourg, France.


\section*{Data Availability Statement}
The catalogs compiled in this work and used throughout this study are provided as supplementary online material. The compilations are also made publicly available online at \url{https://docs.google.com/spreadsheets/d/11b29RREm_rTWcpUGvh7M_-wOQgH8aTiyoxDPHieqXBo/edit?usp=sharing}, in a catalog that we plan to update regularly.


\begin{thebibliography}{}
\expandafter\ifx\csname natexlab\endcsname\relax\def\natexlab#1{#1}\fi
\providecommand{\url}[1]{\href{#1}{#1}}
\providecommand{\dodoi}[1]{doi:~\href{http://doi.org/#1}{\nolinkurl{#1}}}
\providecommand{\doeprint}[1]{\href{http://ascl.net/#1}{\nolinkurl{http://ascl.net/#1}}}
\providecommand{\doarXiv}[1]{\href{https://arxiv.org/abs/#1}{\nolinkurl{https://arxiv.org/abs/#1}}}

\bibitem[{{Adams} {et~al.}(2012){Adams}, {Ciardi}, {Dupree}, {Gautier},
  {Kulesa}, \& {McCarthy}}]{Adams2012}
{Adams}, E.~R., {Ciardi}, D.~R., {Dupree}, A.~K., {et~al.} 2012, \aj, 144, 42,
  \dodoi{10.1088/0004-6256/144/2/42}

\bibitem[{{Adams} {et~al.}(2013){Adams}, {Dupree}, {Kulesa}, \&
  {McCarthy}}]{Adams2013}
{Adams}, E.~R., {Dupree}, A.~K., {Kulesa}, C., \& {McCarthy}, D. 2013, \aj,
  146, 9, \dodoi{10.1088/0004-6256/146/1/9}

\bibitem[{{Artymowicz} \& {Lubow}(1994)}]{ArtymowiczLubow1994}
{Artymowicz}, P., \& {Lubow}, S.~H. 1994, \apj, 421, 651,
  \dodoi{10.1086/173679}

\bibitem[{{Asensio-Torres} {et~al.}(2018){Asensio-Torres}, {Janson},
  {Bonavita}, {Desidera}, {Thalmann}, {Kuzuhara}, {Henning}, {Marzari},
  {Meyer}, {Calissendorff}, \& {Uyama}}]{Asensio-Torres2018}
{Asensio-Torres}, R., {Janson}, M., {Bonavita}, M., {et~al.} 2018, \aap, 619,
  A43, \dodoi{10.1051/0004-6361/201833349}

\bibitem[{{Barnes} {et~al.}(2020){Barnes}, {Haswell}, {Staab},
  {Anglada-Escud{\'e}}, {Fossati}, {Doherty}, {Cooper}, {Jenkins}, {D{\'\i}az},
  {Soto}, \& {Pe{\~n}a Rojas}}]{Barnes2020}
{Barnes}, J.~R., {Haswell}, C.~A., {Staab}, D., {et~al.} 2020, Nature
  Astronomy, 4, 419, \dodoi{10.1038/s41550-019-0972-z}

\bibitem[{{Bate} \& {Bonnell}(1997)}]{BateBonnell1997}
{Bate}, M.~R., \& {Bonnell}, I.~A. 1997, \mnras, 285, 33,
  \dodoi{10.1093/mnras/285.1.33}

\bibitem[{{Bergfors} {et~al.}(2013){Bergfors}, {Brandner}, {Daemgen}, {Biller},
  {Hippler}, {Janson}, {Kudryavtseva}, {Gei{\ss}ler}, {Henning}, \&
  {K{\"o}hler}}]{Bergfors2013}
{Bergfors}, C., {Brandner}, W., {Daemgen}, S., {et~al.} 2013, \mnras, 428, 182,
  \dodoi{10.1093/mnras/sts019}

\bibitem[{{Bochanski} {et~al.}(2010){Bochanski}, {Hawley}, {Covey}, {West},
  {Reid}, {Golimowski}, \& {Ivezi{\'c}}}]{Bochanski2010}
{Bochanski}, J.~J., {Hawley}, S.~L., {Covey}, K.~R., {et~al.} 2010, \aj, 139,
  2679, \dodoi{10.1088/0004-6256/139/6/2679}

\bibitem[{{Bohn} {et~al.}(2020){Bohn}, {Southworth}, {Ginski}, {Kenworthy},
  {Maxted}, \& {Evans}}]{Bohn2020}
{Bohn}, A.~J., {Southworth}, J., {Ginski}, C., {et~al.} 2020, \aap, 635, A73,
  \dodoi{10.1051/0004-6361/201937127}

\bibitem[{{Bonavita} \& {Desidera}(2007)}]{BonavitaDesidera2007}
{Bonavita}, M., \& {Desidera}, S. 2007, \aap, 468, 721,
  \dodoi{10.1051/0004-6361:20066671}

\bibitem[{{Bonavita} \& {Desidera}(2020)}]{BonavitaDesidera2020}
---. 2020, Galaxies, 8, 16, \dodoi{10.3390/galaxies8010016}

\bibitem[{{Bonfils} {et~al.}(2013){Bonfils}, {Delfosse}, {Udry}, {Forveille},
  {Mayor}, {Perrier}, {Bouchy}, {Gillon}, {Lovis}, {Pepe}, {Queloz}, {Santos},
  {S{\'e}gransan}, \& {Bertaux}}]{Bonfils2013}
{Bonfils}, X., {Delfosse}, X., {Udry}, S., {et~al.} 2013, \aap, 549, A109,
  \dodoi{10.1051/0004-6361/201014704}

\bibitem[{{Boss}(2006)}]{Boss2006}
{Boss}, A.~P. 2006, \apj, 641, 1148, \dodoi{10.1086/500530}

\bibitem[{{Chabrier}(2003)}]{Chabrier2003}
{Chabrier}, G. 2003, \pasp, 115, 763, \dodoi{10.1086/376392}

\bibitem[{{Coker} {et~al.}(2018){Coker}, {Gaudi}, {Pogge}, \&
  {Horch}}]{Coker2018}
{Coker}, C.~T., {Gaudi}, B.~S., {Pogge}, R.~W., \& {Horch}, E. 2018, \aj, 155,
  27, \dodoi{10.3847/1538-3881/aa9f0e}

\bibitem[{{Colton} {et~al.}(2021){Colton}, {Horch}, {Everett}, {Howell},
  {Davidson}, {Baptista}, \& {Casetti-Dinescu}}]{Colton2021}
{Colton}, N.~M., {Horch}, E.~P., {Everett}, M.~E., {et~al.} 2021, \aj, 161, 21,
  \dodoi{10.3847/1538-3881/abc9af}

\bibitem[{{Cumming} {et~al.}(2008){Cumming}, {Butler}, {Marcy}, {Vogt},
  {Wright}, \& {Fischer}}]{Cumming2008}
{Cumming}, A., {Butler}, R.~P., {Marcy}, G.~W., {et~al.} 2008, \pasp, 120, 531,
  \dodoi{10.1086/588487}

\bibitem[{{Daemgen} {et~al.}(2009){Daemgen}, {Hormuth}, {Brandner}, {Bergfors},
  {Janson}, {Hippler}, \& {Henning}}]{Daemgen2009}
{Daemgen}, S., {Hormuth}, F., {Brandner}, W., {et~al.} 2009, \aap, 498, 567,
  \dodoi{10.1051/0004-6361/200810988}

\bibitem[{{Deacon} {et~al.}(2014){Deacon}, {Liu}, {Magnier}, {Aller}, {Best},
  {Dupuy}, {Bowler}, {Mann}, {Redstone}, {Burgett}, {Chambers}, {Draper},
  {Flewelling}, {Hodapp}, {Kaiser}, {Kudritzki}, {Morgan}, {Metcalfe}, {Price},
  {Tonry}, \& {Wainscoat}}]{Deacon2014}
{Deacon}, N.~R., {Liu}, M.~C., {Magnier}, E.~A., {et~al.} 2014, \apj, 792, 119,
  \dodoi{10.1088/0004-637X/792/2/119}

\bibitem[{{Deacon} {et~al.}(2016){Deacon}, {Kraus}, {Mann}, {Magnier},
  {Chambers}, {Wainscoat}, {Tonry}, {Kaiser}, {Waters}, {Flewelling}, {Hodapp},
  \& {Burgett}}]{Deacon2016}
{Deacon}, N.~R., {Kraus}, A.~L., {Mann}, A.~W., {et~al.} 2016, \mnras, 455,
  4212, \dodoi{10.1093/mnras/stv2132}

\bibitem[{{Desidera} \& {Barbieri}(2007)}]{DesideraBarbieri2007}
{Desidera}, S., \& {Barbieri}, M. 2007, \aap, 462, 345,
  \dodoi{10.1051/0004-6361:20066319}

\bibitem[{{Dietrich} \& {Ginski}(2018)}]{DietrichGinski2018}
{Dietrich}, J., \& {Ginski}, C. 2018, \aap, 620, A102,
  \dodoi{10.1051/0004-6361/201731341}

\bibitem[{{Dommanget} \& {Nys}(2002)}]{Dommanget2002}
{Dommanget}, J., \& {Nys}, O. 2002, VizieR Online Data Catalog, I/274

\bibitem[{{Doyle} {et~al.}(2011){Doyle}, {Carter}, {Fabrycky}, {Slawson},
  {Howell}, {Winn}, {Orosz}, {P{\v{r}}sa}, {Welsh}, {Quinn}, {Latham},
  {Torres}, {Buchhave}, {Marcy}, {Fortney}, {Shporer}, {Ford}, {Lissauer},
  {Ragozzine}, {Rucker}, {Batalha}, {Jenkins}, {Borucki}, {Koch}, {Middour},
  {Hall}, {McCauliff}, {Fanelli}, {Quintana}, {Holman}, {Caldwell}, {Still},
  {Stefanik}, {Brown}, {Esquerdo}, {Tang}, {Furesz}, {Geary}, {Berlind},
  {Calkins}, {Short}, {Steffen}, {Sasselov}, {Dunham}, {Cochran}, {Boss},
  {Haas}, {Buzasi}, \& {Fischer}}]{Doyle2011}
{Doyle}, L.~R., {Carter}, J.~A., {Fabrycky}, D.~C., {et~al.} 2011, Science,
  333, 1602, \dodoi{10.1126/science.1210923}

\bibitem[{{Eggenberger}(2010)}]{Eggenberger2010}
{Eggenberger}, A. 2010, in EAS Publications Series, Vol.~42, EAS Publications
  Series, ed. K.~{Go{\.z}dziewski}, A.~{Niedzielski}, \& J.~{Schneider},
  19--37, \dodoi{10.1051/eas/1042002}

\bibitem[{{Eggenberger} \& {Udry}(2007)}]{EggenbergerUdry2007}
{Eggenberger}, A., \& {Udry}, S. 2007, arXiv e-prints, arXiv:0705.3173.
\newblock \doarXiv{0705.3173}

\bibitem[{{Eggenberger} {et~al.}(2007){Eggenberger}, {Udry}, {Chauvin},
  {Beuzit}, {Lagrange}, {S{\'e}gransan}, \& {Mayor}}]{Eggenberger2007}
{Eggenberger}, A., {Udry}, S., {Chauvin}, G., {et~al.} 2007, \aap, 474, 273,
  \dodoi{10.1051/0004-6361:20077447}

\bibitem[{{Eggenberger} {et~al.}(2011){Eggenberger}, {Udry}, {Chauvin},
  {Forveille}, {Beuzit}, {Lagrange}, \& {Mayor}}]{Eggenberger2011}
{Eggenberger}, A., {Udry}, S., {Chauvin}, G., {et~al.} 2011, in IAU Symposium,
  Vol. 276, The Astrophysics of Planetary Systems: Formation, Structure, and
  Dynamical Evolution, ed. A.~{Sozzetti}, M.~G. {Lattanzi}, \& A.~P. {Boss},
  409--410, \dodoi{10.1017/S1743921311020564}

\bibitem[{{Eggenberger} {et~al.}(2004){Eggenberger}, {Udry}, \&
  {Mayor}}]{Eggenberger2004}
{Eggenberger}, A., {Udry}, S., \& {Mayor}, M. 2004, \aap, 417, 353,
  \dodoi{10.1051/0004-6361:20034164}

\bibitem[{{Emsenhuber} {et~al.}(2020){Emsenhuber}, {Mordasini}, {Burn},
  {Alibert}, {Benz}, \& {Asphaug}}]{Emsenhuber2020}
{Emsenhuber}, A., {Mordasini}, C., {Burn}, R., {et~al.} 2020, arXiv e-prints,
  arXiv:2007.05562.
\newblock \doarXiv{2007.05562}

\bibitem[{{Everett} {et~al.}(2015){Everett}, {Barclay}, {Ciardi}, {Horch},
  {Howell}, {Crepp}, \& {Silva}}]{Everett2015}
{Everett}, M.~E., {Barclay}, T., {Ciardi}, D.~R., {et~al.} 2015, \aj, 149, 55,
  \dodoi{10.1088/0004-6256/149/2/55}

\bibitem[{{Fabricius} {et~al.}(2001){Fabricius}, {Hog}, {Makarov}, {Mason},
  {Wycoff}, \& {Urban}}]{Fabricius2001}
{Fabricius}, C., {Hog}, E., {Makarov}, V.~V., {et~al.} 2001, VizieR Online Data
  Catalog, I/276

\bibitem[{{Fabrycky} \& {Tremaine}(2007)}]{FabryckyTremaine2007}
{Fabrycky}, D., \& {Tremaine}, S. 2007, \apj, 669, 1298, \dodoi{10.1086/521702}

\bibitem[{{Faedi} {et~al.}(2013){Faedi}, {Staley}, {G{\'o}mez Maqueo Chew},
  {Pollacco}, {Dhital}, {Barros}, {Skillen}, {Hebb}, {Mackay}, \&
  {Watson}}]{Faedi2013}
{Faedi}, F., {Staley}, T., {G{\'o}mez Maqueo Chew}, Y., {et~al.} 2013, \mnras,
  433, 2097, \dodoi{10.1093/mnras/stt885}

\bibitem[{{Fontanive} {et~al.}(2019){Fontanive}, {Rice}, {Bonavita}, {Lopez},
  {Mu{\v{z}}i{\'c}}, \& {Biller}}]{Fontanive2019}
{Fontanive}, C., {Rice}, K., {Bonavita}, M., {et~al.} 2019, \mnras, 485, 4967,
  \dodoi{10.1093/mnras/stz671}

\bibitem[{{Furlan} {et~al.}(2017){Furlan}, {Ciardi}, {Everett}, {Saylors},
  {Teske}, {Horch}, {Howell}, {van Belle}, {Hirsch}, {Gautier}, {Adams},
  {Barrado}, {Cartier}, {Dressing}, {Dupree}, {Gilliland}, {Lillo-Box},
  {Lucas}, \& {Wang}}]{Furlan2017}
{Furlan}, E., {Ciardi}, D.~R., {Everett}, M.~E., {et~al.} 2017, \aj, 153, 71,
  \dodoi{10.3847/1538-3881/153/2/71}

\bibitem[{{Gaia Collaboration} {et~al.}(2016){Gaia Collaboration}, {Prusti},
  {de Bruijne}, {Brown}, {Vallenari}, {Babusiaux}, {Bailer-Jones}, {Bastian},
  {Biermann}, {Evans}, {Eyer}, {Jansen}, {Jordi}, {Klioner}, {Lammers},
  {Lindegren}, {Luri}, {Mignard}, {Milligan}, {Panem}, {Poinsignon},
  {Pourbaix}, {Randich}, {Sarri}, {Sartoretti}, {Siddiqui}, {Soubiran},
  {Valette}, {van Leeuwen}, {Walton}, {Aerts}, {Arenou}, {Cropper}, {Drimmel},
  {H{\o}g}, {Katz}, {Lattanzi}, {O'Mullane}, {Grebel}, {Holland}, {Huc},
  {Passot}, {Bramante}, {Cacciari}, {Casta{\~n}eda}, {Chaoul}, {Cheek}, {De
  Angeli}, {Fabricius}, {Guerra}, {Hern{\'a}ndez}, {Jean-Antoine-Piccolo},
  {Masana}, {Messineo}, {Mowlavi}, {Nienartowicz}, {Ord{\'o}{\~n}ez-Blanco},
  {Panuzzo}, {Portell}, {Richards}, {Riello}, {Seabroke}, {Tanga},
  {Th{\'e}venin}, {Torra}, {Els}, {Gracia-Abril}, {Comoretto},
  {Garcia-Reinaldos}, {Lock}, {Mercier}, {Altmann}, {Andrae}, {Astraatmadja},
  {Bellas-Velidis}, {Benson}, {Berthier}, {Blomme}, {Busso}, {Carry},
  {Cellino}, {Clementini}, {Cowell}, {Creevey}, {Cuypers}, {Davidson}, {De
  Ridder}, {de Torres}, {Delchambre}, {Dell'Oro}, {Ducourant}, {Fr{\'e}mat},
  {Garc{\'\i}a-Torres}, {Gosset}, {Halbwachs}, {Hambly}, {Harrison}, {Hauser},
  {Hestroffer}, {Hodgkin}, {Huckle}, {Hutton}, {Jasniewicz}, {Jordan},
  {Kontizas}, {Korn}, {Lanzafame}, {Manteiga}, {Moitinho}, {Muinonen},
  {Osinde}, {Pancino}, {Pauwels}, {Petit}, {Recio-Blanco}, {Robin}, {Sarro},
  {Siopis}, {Smith}, {Smith}, {Sozzetti}, {Thuillot}, {van Reeven}, {Viala},
  {Abbas}, {Abreu Aramburu}, {Accart}, {Aguado}, {Allan}, {Allasia},
  {Altavilla}, {{\'A}lvarez}, {Alves}, {Anderson}, {Andrei}, {Anglada Varela},
  {Antiche}, {Antoja}, {Ant{\'o}n}, {Arcay}, {Atzei}, {Ayache}, {Bach},
  {Baker}, {Balaguer-N{\'u}{\~n}ez}, {Barache}, {Barata}, {Barbier}, {Barblan},
  {Baroni}, {Barrado y Navascu{\'e}s}, {Barros}, {Barstow}, {Becciani},
  {Bellazzini}, {Bellei}, {Bello Garc{\'\i}a}, {Belokurov}, {Bendjoya},
  {Berihuete}, {Bianchi}, {Bienaym{\'e}}, {Billebaud}, {Blagorodnova},
  {Blanco-Cuaresma}, {Boch}, {Bombrun}, {Borrachero}, {Bouquillon}, {Bourda},
  {Bouy}, {Bragaglia}, {Breddels}, {Brouillet}, {Br{\"u}semeister},
  {Bucciarelli}, {Budnik}, {Burgess}, {Burgon}, {Burlacu}, {Busonero}, {Buzzi},
  {Caffau}, {Cambras}, {Campbell}, {Cancelliere}, {Cantat-Gaudin}, {Carlucci},
  {Carrasco}, {Castellani}, {Charlot}, {Charnas}, {Charvet}, {Chassat},
  {Chiavassa}, {Clotet}, {Cocozza}, {Collins}, {Collins}, {Costigan}, {Crifo},
  {Cross}, {Crosta}, {Crowley}, {Dafonte}, {Damerdji}, {Dapergolas}, {David},
  {David}, {De Cat}, {de Felice}, {de Laverny}, {De Luise}, {De March}, {de
  Martino}, {de Souza}, {Debosscher}, {del Pozo}, {Delbo}, {Delgado},
  {Delgado}, {di Marco}, {Di Matteo}, {Diakite}, {Distefano}, {Dolding}, {Dos
  Anjos}, {Drazinos}, {Dur{\'a}n}, {Dzigan}, {Ecale}, {Edvardsson}, {Enke},
  {Erdmann}, {Escolar}, {Espina}, {Evans}, {Eynard Bontemps}, {Fabre},
  {Fabrizio}, {Faigler}, {Falc{\~a}o}, {Farr{\`a}s Casas}, {Faye}, {Federici},
  {Fedorets}, {Fern{\'a}ndez-Hern{\'a}ndez}, {Fernique}, {Fienga}, {Figueras},
  {Filippi}, {Findeisen}, {Fonti}, {Fouesneau}, {Fraile}, {Fraser}, {Fuchs},
  {Furnell}, {Gai}, {Galleti}, {Galluccio}, {Garabato}, {Garc{\'\i}a-Sedano},
  {Gar{\'e}}, {Garofalo}, {Garralda}, {Gavras}, {Gerssen}, {Geyer}, {Gilmore},
  {Girona}, {Giuffrida}, {Gomes}, {Gonz{\'a}lez-Marcos},
  {Gonz{\'a}lez-N{\'u}{\~n}ez}, {Gonz{\'a}lez-Vidal}, {Granvik}, {Guerrier},
  {Guillout}, {Guiraud}, {G{\'u}rpide}, {Guti{\'e}rrez-S{\'a}nchez}, {Guy},
  {Haigron}, {Hatzidimitriou}, {Haywood}, {Heiter}, {Helmi}, {Hobbs},
  {Hofmann}, {Holl}, {Holland }, {Hunt}, {Hypki}, {Icardi}, {Irwin}, {Jevardat
  de Fombelle}, {Jofr{\'e}}, {Jonker}, {Jorissen}, {Julbe}, {Karampelas},
  {Kochoska}, {Kohley}, {Kolenberg}, {Kontizas}, {Koposov}, {Kordopatis},
  {Koubsky}, {Kowalczyk}, {Krone-Martins}, {Kudryashova}, {Kull}, {Bachchan},
  {Lacoste-Seris}, {Lanza}, {Lavigne}, {Le Poncin-Lafitte}, {Lebreton},
  {Lebzelter}, {Leccia}, {Leclerc}, {Lecoeur-Taibi}, {Lemaitre}, {Lenhardt},
  {Leroux}, {Liao}, {Licata}, {Lindstr{\o}m}, {Lister}, {Livanou}, {Lobel},
  {L{\"o}ffler}, {L{\'o}pez}, {Lopez-Lozano}, {Lorenz}, {Loureiro},
  {MacDonald}, {Magalh{\~a}es Fernandes}, {Managau}, {Mann}, {Mantelet},
  {Marchal}, {Marchant}, {Marconi}, {Marie}, {Marinoni}, {Marrese},
  {Marschalk{\'o}}, {Marshall}, {Mart{\'\i}n-Fleitas}, {Martino}, {Mary},
  {Matijevi{\v{c}}}, {Mazeh}, {McMillan}, {Messina}, {Mestre}, {Michalik},
  {Millar}, {Miranda}, {Molina}, {Molinaro}, {Molinaro}, {Moln{\'a}r},
  {Moniez}, {Montegriffo}, {Monteiro}, {Mor}, {Mora}, {Morbidelli}, {Morel},
  {Morgenthaler}, {Morley}, {Morris}, {Mulone}, {Muraveva}, {Musella},
  {Narbonne}, {Nelemans}, {Nicastro}, {Noval}, {Ord{\'e}novic},
  {Ordieres-Mer{\'e}}, {Osborne}, {Pagani}, {Pagano}, {Pailler}, {Palacin},
  {Palaversa}, {Parsons}, {Paulsen}, {Pecoraro}, {Pedrosa}, {Pentik{\"a}inen},
  {Pereira}, {Pichon}, {Piersimoni}, {Pineau}, {Plachy}, {Plum}, {Poujoulet},
  {Pr{\v{s}}a}, {Pulone}, {Ragaini}, {Rago}, {Rambaux}, {Ramos-Lerate},
  {Ranalli}, {Rauw}, {Read}, {Regibo}, {Renk}, {Reyl{\'e}}, {Ribeiro},
  {Rimoldini}, {Ripepi}, {Riva}, {Rixon}, {Roelens}, {Romero-G{\'o}mez},
  {Rowell}, {Royer}, {Rudolph}, {Ruiz-Dern}, {Sadowski}, {Sagrist{\`a}
  Sell{\'e}s}, {Sahlmann}, {Salgado}, {Salguero}, {Sarasso}, {Savietto},
  {Schnorhk}, {Schultheis}, {Sciacca}, {Segol}, {Segovia}, {Segransan},
  {Serpell}, {Shih}, {Smareglia}, {Smart}, {Smith}, {Solano}, {Solitro},
  {Sordo}, {Soria Nieto}, {Souchay}, {Spagna}, {Spoto}, {Stampa}, {Steele},
  {Steidelm{\"u}ller}, {Stephenson}, {Stoev}, {Suess}, {S{\"u}veges}, {Surdej},
  {Szabados}, {Szegedi-Elek}, {Tapiador}, {Taris}, {Tauran}, {Taylor},
  {Teixeira}, {Terrett}, {Tingley}, {Trager}, {Turon}, {Ulla}, {Utrilla},
  {Valentini}, {van Elteren}, {Van Hemelryck}, {van Leeuwen}, {Varadi},
  {Vecchiato}, {Veljanoski}, {Via}, {Vicente}, {Vogt}, {Voss}, {Votruba},
  {Voutsinas}, {Walmsley}, {Weiler}, {Weingrill}, {Werner}, {Wevers},
  {Whitehead}, {Wyrzykowski}, {Yoldas}, {{\v{Z}}erjal}, {Zucker}, {Zurbach},
  {Zwitter}, {Alecu}, {Allen}, {Allende Prieto}, {Amorim},
  {Anglada-Escud{\'e}}, {Arsenijevic}, {Azaz}, {Balm}, {Beck}, {Bernstein},
  {Bigot}, {Bijaoui}, {Blasco}, {Bonfigli}, {Bono}, {Boudreault}, {Bressan},
  {Brown}, {Brunet}, {Bunclark}, {Buonanno}, {Butkevich}, {Carret}, {Carrion},
  {Chemin}, {Ch{\'e}reau}, {Corcione}, {Darmigny}, {de Boer}, {de Teodoro}, {de
  Zeeuw}, {Delle Luche}, {Domingues}, {Dubath}, {Fodor}, {Fr{\'e}zouls},
  {Fries}, {Fustes}, {Fyfe}, {Gallardo}, {Gallegos}, {Gardiol}, {Gebran},
  {Gomboc}, {G{\'o}mez}, {Grux}, {Gueguen}, {Heyrovsky}, {Hoar}, {Iannicola},
  {Isasi Parache}, {Janotto}, {Joliet}, {Jonckheere}, {Keil}, {Kim},
  {Klagyivik}, {Klar}, {Knude}, {Kochukhov}, {Kolka}, {Kos}, {Kutka}, {Lainey},
  {LeBouquin}, {Liu}, {Loreggia}, {Makarov}, {Marseille}, {Martayan},
  {Martinez-Rubi}, {Massart}, {Meynadier}, {Mignot}, {Munari}, {Nguyen},
  {Nordlander}, {Ocvirk}, {O'Flaherty}, {Olias Sanz}, {Ortiz}, {Osorio},
  {Oszkiewicz}, {Ouzounis}, {Palmer}, {Park}, {Pasquato}, {Peltzer}, {Peralta},
  {P{\'e}turaud}, {Pieniluoma}, {Pigozzi}, {Poels}, {Prat}, {Prod'homme},
  {Raison}, {Rebordao}, {Risquez}, {Rocca-Volmerange}, {Rosen}, {Ruiz-Fuertes},
  {Russo}, {Sembay}, {Serraller Vizcaino}, {Short}, {Siebert}, {Silva},
  {Sinachopoulos}, {Slezak}, {Soffel}, {Sosnowska}, {Strai{\v{z}}ys}, {ter
  Linden}, {Terrell}, {Theil}, {Tiede}, {Troisi}, {Tsalmantza}, {Tur},
  {Vaccari}, {Vachier}, {Valles}, {Van Hamme}, {Veltz}, {Virtanen}, {Wallut},
  {Wichmann}, {Wilkinson}, {Ziaeepour}, \& {Zschocke}}]{Gaia2016}
{Gaia Collaboration}, {Prusti}, T., {de Bruijne}, J.~H.~J., {et~al.} 2016,
  \aap, 595, A1, \dodoi{10.1051/0004-6361/201629272}

\bibitem[{{Gaia Collaboration} {et~al.}(2018){Gaia Collaboration}, {Brown},
  {Vallenari}, {Prusti}, {de Bruijne}, {Babusiaux}, {Bailer-Jones}, {Biermann},
  {Evans}, {Eyer}, {Jansen}, {Jordi}, {Klioner}, {Lammers}, {Lindegren},
  {Luri}, {Mignard}, {Panem}, {Pourbaix}, {Randich}, {Sartoretti}, {Siddiqui},
  {Soubiran}, {van Leeuwen}, {Walton}, {Arenou}, {Bastian}, {Cropper},
  {Drimmel}, {Katz}, {Lattanzi}, {Bakker}, {Cacciari}, {Casta{\~n}eda},
  {Chaoul}, {Cheek}, {De Angeli}, {Fabricius}, {Guerra}, {Holl}, {Masana},
  {Messineo}, {Mowlavi}, {Nienartowicz}, {Panuzzo}, {Portell}, {Riello},
  {Seabroke}, {Tanga}, {Th{\'e}venin}, {Gracia-Abril}, {Comoretto},
  {Garcia-Reinaldos}, {Teyssier}, {Altmann}, {Andrae}, {Audard},
  {Bellas-Velidis}, {Benson}, {Berthier}, {Blomme}, {Burgess}, {Busso},
  {Carry}, {Cellino}, {Clementini}, {Clotet}, {Creevey}, {Davidson}, {De
  Ridder}, {Delchambre}, {Dell'Oro}, {Ducourant},
  {Fern{\'a}ndez-Hern{\'a}ndez}, {Fouesneau}, {Fr{\'e}mat}, {Galluccio},
  {Garc{\'\i}a-Torres}, {Gonz{\'a}lez-N{\'u}{\~n}ez}, {Gonz{\'a}lez-Vidal},
  {Gosset}, {Guy}, {Halbwachs}, {Hambly}, {Harrison}, {Hern{\'a}ndez},
  {Hestroffer}, {Hodgkin}, {Hutton}, {Jasniewicz}, {Jean-Antoine-Piccolo},
  {Jordan}, {Korn}, {Krone-Martins}, {Lanzafame}, {Lebzelter}, {L{\"o}ffler},
  {Manteiga}, {Marrese}, {Mart{\'\i}n-Fleitas}, {Moitinho}, {Mora}, {Muinonen},
  {Osinde}, {Pancino}, {Pauwels}, {Petit}, {Recio-Blanco}, {Richards},
  {Rimoldini}, {Robin}, {Sarro}, {Siopis}, {Smith}, {Sozzetti}, {S{\"u}veges},
  {Torra}, {van Reeven}, {Abbas}, {Abreu Aramburu}, {Accart}, {Aerts},
  {Altavilla}, {{\'A}lvarez}, {Alvarez}, {Alves}, {Anderson}, {Andrei},
  {Anglada Varela}, {Antiche}, {Antoja}, {Arcay}, {Astraatmadja}, {Bach},
  {Baker}, {Balaguer-N{\'u}{\~n}ez}, {Balm}, {Barache}, {Barata}, {Barbato},
  {Barblan}, {Barklem}, {Barrado}, {Barros}, {Barstow}, {Bartholom{\'e}
  Mu{\~n}oz}, {Bassilana}, {Becciani}, {Bellazzini}, {Berihuete}, {Bertone},
  {Bianchi}, {Bienaym{\'e}}, {Blanco-Cuaresma}, {Boch}, {Boeche}, {Bombrun},
  {Borrachero}, {Bossini}, {Bouquillon}, {Bourda}, {Bragaglia}, {Bramante},
  {Breddels}, {Bressan}, {Brouillet}, {Br{\"u}semeister}, {Brugaletta},
  {Bucciarelli}, {Burlacu}, {Busonero}, {Butkevich}, {Buzzi}, {Caffau},
  {Cancelliere}, {Cannizzaro}, {Cantat-Gaudin}, {Carballo}, {Carlucci},
  {Carrasco}, {Casamiquela}, {Castellani}, {Castro-Ginard}, {Charlot},
  {Chemin}, {Chiavassa}, {Cocozza}, {Costigan}, {Cowell}, {Crifo}, {Crosta},
  {Crowley}, {Cuypers}, {Dafonte}, {Damerdji}, {Dapergolas}, {David}, {David},
  {de Laverny}, {De Luise}, {De March}, {de Martino}, {de Souza}, {de Torres},
  {Debosscher}, {del Pozo}, {Delbo}, {Delgado}, {Delgado}, {Di Matteo},
  {Diakite}, {Diener}, {Distefano}, {Dolding}, {Drazinos}, {Dur{\'a}n},
  {Edvardsson}, {Enke}, {Eriksson}, {Esquej}, {Eynard Bontemps}, {Fabre},
  {Fabrizio}, {Faigler}, {Falc{\~a}o}, {Farr{\`a}s Casas}, {Federici},
  {Fedorets}, {Fernique}, {Figueras}, {Filippi}, {Findeisen}, {Fonti},
  {Fraile}, {Fraser}, {Fr{\'e}zouls}, {Gai}, {Galleti}, {Garabato},
  {Garc{\'\i}a-Sedano}, {Garofalo}, {Garralda}, {Gavel}, {Gavras}, {Gerssen},
  {Geyer}, {Giacobbe}, {Gilmore}, {Girona}, {Giuffrida}, {Glass}, {Gomes},
  {Granvik}, {Gueguen}, {Guerrier}, {Guiraud}, {Guti{\'e}rrez-S{\'a}nchez},
  {Haigron}, {Hatzidimitriou}, {Hauser}, {Haywood}, {Heiter}, {Helmi}, {Heu},
  {Hilger}, {Hobbs}, {Hofmann}, {Holland}, {Huckle}, {Hypki}, {Icardi},
  {Jan{\ss}en}, {Jevardat de Fombelle}, {Jonker}, {Juh{\'a}sz}, {Julbe},
  {Karampelas}, {Kewley}, {Klar}, {Kochoska}, {Kohley}, {Kolenberg},
  {Kontizas}, {Kontizas}, {Koposov}, {Kordopatis}, {Kostrzewa-Rutkowska},
  {Koubsky}, {Lambert}, {Lanza}, {Lasne}, {Lavigne}, {Le Fustec}, {Le
  Poncin-Lafitte}, {Lebreton}, {Leccia}, {Leclerc}, {Lecoeur-Taibi},
  {Lenhardt}, {Leroux}, {Liao}, {Licata}, {Lindstr{\o}m}, {Lister}, {Livanou},
  {Lobel}, {L{\'o}pez}, {Managau}, {Mann}, {Mantelet}, {Marchal}, {Marchant},
  {Marconi}, {Marinoni}, {Marschalk{\'o}}, {Marshall}, {Martino}, {Marton},
  {Mary}, {Massari}, {Matijevi{\v{c}}}, {Mazeh}, {McMillan}, {Messina},
  {Michalik}, {Millar}, {Molina}, {Molinaro}, {Moln{\'a}r}, {Montegriffo},
  {Mor}, {Morbidelli}, {Morel}, {Morris}, {Mulone}, {Muraveva}, {Musella},
  {Nelemans}, {Nicastro}, {Noval}, {O'Mullane}, {Ord{\'e}novic},
  {Ord{\'o}{\~n}ez-Blanco}, {Osborne}, {Pagani}, {Pagano}, {Pailler},
  {Palacin}, {Palaversa}, {Panahi}, {Pawlak}, {Piersimoni}, {Pineau}, {Plachy},
  {Plum}, {Poggio}, {Poujoulet}, {Pr{\v{s}}a}, {Pulone}, {Racero}, {Ragaini},
  {Rambaux}, {Ramos-Lerate}, {Regibo}, {Reyl{\'e}}, {Riclet}, {Ripepi}, {Riva},
  {Rivard}, {Rixon}, {Roegiers}, {Roelens}, {Romero-G{\'o}mez}, {Rowell},
  {Royer}, {Ruiz-Dern}, {Sadowski}, {Sagrist{\`a} Sell{\'e}s}, {Sahlmann},
  {Salgado}, {Salguero}, {Sanna}, {Santana-Ros}, {Sarasso}, {Savietto},
  {Schultheis}, {Sciacca}, {Segol}, {Segovia}, {S{\'e}gransan}, {Shih},
  {Siltala}, {Silva}, {Smart}, {Smith}, {Solano}, {Solitro}, {Sordo}, {Soria
  Nieto}, {Souchay}, {Spagna}, {Spoto}, {Stampa}, {Steele},
  {Steidelm{\"u}ller}, {Stephenson}, {Stoev}, {Suess}, {Surdej}, {Szabados},
  {Szegedi-Elek}, {Tapiador}, {Taris}, {Tauran}, {Taylor}, {Teixeira},
  {Terrett}, {Teyssand ier}, {Thuillot}, {Titarenko}, {Torra Clotet}, {Turon},
  {Ulla}, {Utrilla}, {Uzzi}, {Vaillant}, {Valentini}, {Valette}, {van Elteren},
  {Van Hemelryck}, {van Leeuwen}, {Vaschetto}, {Vecchiato}, {Veljanoski},
  {Viala}, {Vicente}, {Vogt}, {von Essen}, {Voss}, {Votruba}, {Voutsinas},
  {Walmsley}, {Weiler}, {Wertz}, {Wevers}, {Wyrzykowski}, {Yoldas},
  {{\v{Z}}erjal}, {Ziaeepour}, {Zorec}, {Zschocke}, {Zucker}, {Zurbach}, \&
  {Zwitter}}]{Gaia2018}
{Gaia Collaboration}, {Brown}, A.~G.~A., {Vallenari}, A., {et~al.} 2018, \aap,
  616, A1, \dodoi{10.1051/0004-6361/201833051}

\bibitem[{{Gentile Fusillo} {et~al.}(2019){Gentile Fusillo}, {Tremblay},
  {G{\"a}nsicke}, {Manser}, {Cunningham}, {Cukanovaite}, {Hollands}, {Marsh},
  {Raddi}, {Jordan}, {Toonen}, {Geier}, {Barstow}, \&
  {Cummings}}]{GentileFusillo2019}
{Gentile Fusillo}, N.~P., {Tremblay}, P.-E., {G{\"a}nsicke}, B.~T., {et~al.}
  2019, \mnras, 482, 4570, \dodoi{10.1093/mnras/sty3016}

\bibitem[{{Ginski} {et~al.}(2020){Ginski}, {Mugrauer}, {Adam}, {Vogt}, \& {van
  Holstein}}]{Ginski2020}
{Ginski}, C., {Mugrauer}, M., {Adam}, C., {Vogt}, N., \& {van Holstein}, R.
  2020, arXiv e-prints, arXiv:2009.10363.
\newblock \doarXiv{2009.10363}

\bibitem[{{Ginski} {et~al.}(2012){Ginski}, {Mugrauer}, {Seeliger}, \&
  {Eisenbeiss}}]{Ginski2012}
{Ginski}, C., {Mugrauer}, M., {Seeliger}, M., \& {Eisenbeiss}, T. 2012, \mnras,
  421, 2498, \dodoi{10.1111/j.1365-2966.2012.20485.x}

\bibitem[{{Ginski} {et~al.}(2016){Ginski}, {Mugrauer}, {Seeliger}, {Buder},
  {Errmann}, {Avenhaus}, {Mouillet}, {Maire}, \& {Raetz}}]{Ginski2016}
{Ginski}, C., {Mugrauer}, M., {Seeliger}, M., {et~al.} 2016, \mnras, 457, 2173,
  \dodoi{10.1093/mnras/stw049}

\bibitem[{{Hagelberg} {et~al.}(2020){Hagelberg}, {Engler}, {Fontanive},
  {Daemgen}, {Quanz}, {K{\"u}hn}, {Reggiani}, {Meyer}, {Jayawardhana}, \&
  {Kostov}}]{Hagelberg2020}
{Hagelberg}, J., {Engler}, N., {Fontanive}, C., {et~al.} 2020, \aap, 643, A98,
  \dodoi{10.1051/0004-6361/202039173}

\bibitem[{{Hall} {et~al.}(2017){Hall}, {Forgan}, \& {Rice}}]{Hall2017}
{Hall}, C., {Forgan}, D., \& {Rice}, K. 2017, \mnras, 470, 2517,
  \dodoi{10.1093/mnras/stx1244}

\bibitem[{{Han} {et~al.}(2014){Han}, {Wang}, {Wright}, {Feng}, {Zhao},
  {Fakhouri}, {Brown}, \& {Hancock}}]{Han2014}
{Han}, E., {Wang}, S.~X., {Wright}, J.~T., {et~al.} 2014, \pasp, 126, 827,
  \dodoi{10.1086/678447}

\bibitem[{{Hatzes} {et~al.}(2003){Hatzes}, {Cochran}, {Endl}, {McArthur},
  {Paulson}, {Walker}, {Campbell}, \& {Yang}}]{Hatzes2003}
{Hatzes}, A.~P., {Cochran}, W.~D., {Endl}, M., {et~al.} 2003, \apj, 599, 1383,
  \dodoi{10.1086/379281}

\bibitem[{{Hirsch} {et~al.}(2017){Hirsch}, {Ciardi}, {Howard}, {Everett},
  {Furlan}, {Saylors}, {Horch}, {Howell}, {Teske}, \& {Marcy}}]{Hirsch2017}
{Hirsch}, L.~A., {Ciardi}, D.~R., {Howard}, A.~W., {et~al.} 2017, \aj, 153,
  117, \dodoi{10.3847/1538-3881/153/3/117}

\bibitem[{{Horch} {et~al.}(2014){Horch}, {Howell}, {Everett}, \&
  {Ciardi}}]{Horch2014}
{Horch}, E.~P., {Howell}, S.~B., {Everett}, M.~E., \& {Ciardi}, D.~R. 2014,
  \apj, 795, 60, \dodoi{10.1088/0004-637X/795/1/60}

\bibitem[{{Howard} {et~al.}(2012){Howard}, {Marcy}, {Bryson}, {Jenkins},
  {Rowe}, {Batalha}, {Borucki}, {Koch}, {Dunham}, {Gautier}, {Van Cleve},
  {Cochran}, {Latham}, {Lissauer}, {Torres}, {Brown}, {Gilliland}, {Buchhave},
  {Caldwell}, {Christensen-Dalsgaard}, {Ciardi}, {Fressin}, {Haas}, {Howell},
  {Kjeldsen}, {Seager}, {Rogers}, {Sasselov}, {Steffen}, {Basri},
  {Charbonneau}, {Christiansen}, {Clarke}, {Dupree}, {Fabrycky}, {Fischer},
  {Ford}, {Fortney}, {Tarter}, {Girouard}, {Holman}, {Johnson}, {Klaus},
  {Machalek}, {Moorhead}, {Morehead}, {Ragozzine}, {Tenenbaum}, {Twicken},
  {Quinn}, {Isaacson}, {Shporer}, {Lucas}, {Walkowicz}, {Welsh}, {Boss},
  {Devore}, {Gould}, {Smith}, {Morris}, {Prsa}, {Morton}, {Still}, {Thompson},
  {Mullally}, {Endl}, \& {MacQueen}}]{Howard2012}
{Howard}, A.~W., {Marcy}, G.~W., {Bryson}, S.~T., {et~al.} 2012, \apjs, 201,
  15, \dodoi{10.1088/0067-0049/201/2/15}

\bibitem[{{Janson} {et~al.}(2014){Janson}, {Bergfors}, {Brandner},
  {Kudryavtseva}, {Hormuth}, {Hippler}, \& {Henning}}]{Janson2014}
{Janson}, M., {Bergfors}, C., {Brandner}, W., {et~al.} 2014, \apj, 789, 102,
  \dodoi{10.1088/0004-637X/789/2/102}

\bibitem[{{Janson} {et~al.}(2017){Janson}, {Durkan}, {Hippler}, {Dai},
  {Brandner}, {Schlieder}, {Bonnefoy}, \& {Henning}}]{Janson2017}
{Janson}, M., {Durkan}, S., {Hippler}, S., {et~al.} 2017, \aap, 599, A70,
  \dodoi{10.1051/0004-6361/201629945}

\bibitem[{{Janson} {et~al.}(2012){Janson}, {Hormuth}, {Bergfors}, {Brand ner},
  {Hippler}, {Daemgen}, {Kudryavtseva}, {Schmalzl}, {Schnupp}, \&
  {Henning}}]{Janson2012}
{Janson}, M., {Hormuth}, F., {Bergfors}, C., {et~al.} 2012, \apj, 754, 44,
  \dodoi{10.1088/0004-637X/754/1/44}

\bibitem[{{Jensen} \& {Akeson}(2003)}]{JensenAkeson2003}
{Jensen}, E. L.~N., \& {Akeson}, R.~L. 2003, \apj, 584, 875,
  \dodoi{10.1086/345719}

\bibitem[{{Kaib} {et~al.}(2013){Kaib}, {Raymond}, \& {Duncan}}]{Kaib2013}
{Kaib}, N.~A., {Raymond}, S.~N., \& {Duncan}, M. 2013, \nat, 493, 381,
  \dodoi{10.1038/nature11780}

\bibitem[{{Kley}(2001)}]{Kley2001}
{Kley}, W. 2001, in The Formation of Binary Stars, ed. H.~{Zinnecker} \&
  R.~{Mathieu}, Vol. 200, 511

\bibitem[{{Konacki} {et~al.}(2009){Konacki}, {Muterspaugh}, {Kulkarni}, \&
  {He{\l}miniak}}]{Konacki2009}
{Konacki}, M., {Muterspaugh}, M.~W., {Kulkarni}, S.~R., \& {He{\l}miniak},
  K.~G. 2009, \apj, 704, 513, \dodoi{10.1088/0004-637X/704/1/513}

\bibitem[{{Kouwenhoven} {et~al.}(2007){Kouwenhoven}, {Brown}, {Portegies
  Zwart}, \& {Kaper}}]{Kouwenhoven2007}
{Kouwenhoven}, M.~B.~N., {Brown}, A.~G.~A., {Portegies Zwart}, S.~F., \&
  {Kaper}, L. 2007, \aap, 474, 77, \dodoi{10.1051/0004-6361:20077719}

\bibitem[{{Kozai}(1962)}]{Kozai1962}
{Kozai}, Y. 1962, \aj, 67, 591, \dodoi{10.1086/108790}

\bibitem[{{Kraus} {et~al.}(2012){Kraus}, {Ireland}, {Hillenbrand}, \&
  {Martinache}}]{Kraus2012}
{Kraus}, A.~L., {Ireland}, M.~J., {Hillenbrand}, L.~A., \& {Martinache}, F.
  2012, \apj, 745, 19, \dodoi{10.1088/0004-637X/745/1/19}

\bibitem[{{Kraus} {et~al.}(2016){Kraus}, {Ireland}, {Huber}, {Mann}, \&
  {Dupuy}}]{Kraus2016}
{Kraus}, A.~L., {Ireland}, M.~J., {Huber}, D., {Mann}, A.~W., \& {Dupuy}, T.~J.
  2016, \aj, 152, 8, \dodoi{10.3847/0004-6256/152/1/8}

\bibitem[{{Lidov}(1962)}]{Lidov1962}
{Lidov}, M.~L. 1962, \planss, 9, 719, \dodoi{10.1016/0032-0633(62)90129-0}

\bibitem[{{Lillo-Box} {et~al.}(2012){Lillo-Box}, {Barrado}, \&
  {Bouy}}]{Lillo-Box2012}
{Lillo-Box}, J., {Barrado}, D., \& {Bouy}, H. 2012, \aap, 546, A10,
  \dodoi{10.1051/0004-6361/201219631}

\bibitem[{{Lodieu} {et~al.}(2014){Lodieu}, {P{\'e}rez-Garrido}, {B{\'e}jar},
  {Gauza}, {Ruiz}, {Rebolo}, {Pinfield}, \& {Mart{\'\i}n}}]{Lodieu2014}
{Lodieu}, N., {P{\'e}rez-Garrido}, A., {B{\'e}jar}, V.~J.~S., {et~al.} 2014,
  \aap, 569, A120, \dodoi{10.1051/0004-6361/201424210}

\bibitem[{{Luhman} \& {Jayawardhana}(2002)}]{LuhmanJayawardhana2002}
{Luhman}, K.~L., \& {Jayawardhana}, R. 2002, \apj, 566, 1132,
  \dodoi{10.1086/338338}

\bibitem[{{Maldonado} {et~al.}(2020){Maldonado}, {Villaver}, {Mustill},
  {Chavez}, \& {Bertone}}]{Maldonado2020}
{Maldonado}, R.~F., {Villaver}, E., {Mustill}, A.~J., {Chavez}, M., \&
  {Bertone}, E. 2020, \mnras, 499, 1854, \dodoi{10.1093/mnras/staa2946}

\bibitem[{{Mason} {et~al.}(2001){Mason}, {Wycoff}, {Hartkopf}, {Douglass}, \&
  {Worley}}]{Mason2001}
{Mason}, B.~D., {Wycoff}, G.~L., {Hartkopf}, W.~I., {Douglass}, G.~G., \&
  {Worley}, C.~E. 2001, \aj, 122, 3466, \dodoi{10.1086/323920}

\bibitem[{{Matson} {et~al.}(2018){Matson}, {Howell}, {Horch}, \&
  {Everett}}]{Matson2018}
{Matson}, R.~A., {Howell}, S.~B., {Horch}, E.~P., \& {Everett}, M.~E. 2018,
  \aj, 156, 31, \dodoi{10.3847/1538-3881/aac778}

\bibitem[{{Mayor} {et~al.}(2011){Mayor}, {Marmier}, {Lovis}, {Udry},
  {S{\'e}gransan}, {Pepe}, {Benz}, {Bertaux}, {Bouchy}, {Dumusque}, {Lo Curto},
  {Mordasini}, {Queloz}, \& {Santos}}]{Mayor2011}
{Mayor}, M., {Marmier}, M., {Lovis}, C., {et~al.} 2011, arXiv e-prints,
  arXiv:1109.2497.
\newblock \doarXiv{1109.2497}

\bibitem[{{Moe} \& {Kratter}(2018)}]{MoeKratter2018}
{Moe}, M., \& {Kratter}, K.~M. 2018, \apj, 854, 44,
  \dodoi{10.3847/1538-4357/aaa6d2}

\bibitem[{{Moe} \& {Kratter}(2019)}]{MoeKratter2019}
---. 2019, arXiv e-prints, arXiv:1912.01699.
\newblock \doarXiv{1912.01699}

\bibitem[{{Mordasini}(2018)}]{Mordasini2018}
{Mordasini}, C. 2018, {Planetary Population Synthesis}, ed. H.~J. {Deeg} \&
  J.~A. {Belmonte}, 143, \dodoi{10.1007/978-3-319-55333-7_143}

\bibitem[{{Moutou} {et~al.}(2017){Moutou}, {Vigan}, {Mesa}, {Desidera},
  {Th{\'e}bault}, {Zurlo}, \& {Salter}}]{Moutou2017}
{Moutou}, C., {Vigan}, A., {Mesa}, D., {et~al.} 2017, \aap, 602, A87,
  \dodoi{10.1051/0004-6361/201630173}

\bibitem[{{Mugrauer}(2019)}]{Mugrauer2019}
{Mugrauer}, M. 2019, \mnras, 490, 5088, \dodoi{10.1093/mnras/stz2673}

\bibitem[{{Mugrauer} \& {Ginski}(2015)}]{MugrauerGinski2015}
{Mugrauer}, M., \& {Ginski}, C. 2015, \mnras, 450, 3127,
  \dodoi{10.1093/mnras/stv771}

\bibitem[{{Mugrauer} \& {Neuh{\"a}user}(2009)}]{MugrauerNeuhauser2009}
{Mugrauer}, M., \& {Neuh{\"a}user}, R. 2009, \aap, 494, 373,
  \dodoi{10.1051/0004-6361:200810639}

\bibitem[{{Mugrauer} {et~al.}(2007{\natexlab{a}}){Mugrauer}, {Neuh{\"a}user},
  \& {Mazeh}}]{Mugrauer2007b}
{Mugrauer}, M., {Neuh{\"a}user}, R., \& {Mazeh}, T. 2007{\natexlab{a}}, \aap,
  469, 755, \dodoi{10.1051/0004-6361:20065883}

\bibitem[{{Mugrauer} {et~al.}(2006){Mugrauer}, {Neuh{\"a}user}, {Mazeh},
  {Guenther}, {Fern{\'a}ndez}, \& {Broeg}}]{Mugrauer2006}
{Mugrauer}, M., {Neuh{\"a}user}, R., {Mazeh}, T., {et~al.} 2006, Astronomische
  Nachrichten, 327, 321, \dodoi{10.1002/asna.200510528}

\bibitem[{{Mugrauer} {et~al.}(2007{\natexlab{b}}){Mugrauer}, {Seifahrt}, \&
  {Neuh{\"a}user}}]{Mugrauer2007}
{Mugrauer}, M., {Seifahrt}, A., \& {Neuh{\"a}user}, R. 2007{\natexlab{b}},
  \mnras, 378, 1328, \dodoi{10.1111/j.1365-2966.2007.11858.x}

\bibitem[{{Mulders} {et~al.}(2015){Mulders}, {Pascucci}, \&
  {Apai}}]{Mulders2015}
{Mulders}, G.~D., {Pascucci}, I., \& {Apai}, D. 2015, \apj, 798, 112,
  \dodoi{10.1088/0004-637X/798/2/112}

\bibitem[{{M{\"u}ller} \& {Kley}(2012)}]{Muller2012}
{M{\"u}ller}, T.~W.~A., \& {Kley}, W. 2012, \aap, 539, A18,
  \dodoi{10.1051/0004-6361/201118202}

\bibitem[{{Naoz} \& {Fabrycky}(2014)}]{NaozFabrycky2014}
{Naoz}, S., \& {Fabrycky}, D.~C. 2014, \apj, 793, 137,
  \dodoi{10.1088/0004-637X/793/2/137}

\bibitem[{{Ngo} {et~al.}(2016){Ngo}, {Knutson}, {Hinkley}, {Bryan}, {Crepp},
  {Batygin}, {Crossfield}, {Hansen}, {Howard}, {Johnson}, {Mawet}, {Morton},
  {Muirhead}, \& {Wang}}]{Ngo2016}
{Ngo}, H., {Knutson}, H.~A., {Hinkley}, S., {et~al.} 2016, \apj, 827, 8,
  \dodoi{10.3847/0004-637X/827/1/8}

\bibitem[{{Ngo} {et~al.}(2017){Ngo}, {Knutson}, {Bryan}, {Blunt}, {Nielsen},
  {Batygin}, {Bowler}, {Crepp}, {Hinkley}, {Howard}, \& {Mawet}}]{Ngo2017}
{Ngo}, H., {Knutson}, H.~A., {Bryan}, M.~L., {et~al.} 2017, \aj, 153, 242,
  \dodoi{10.3847/1538-3881/aa6cac}

\bibitem[{{Ortiz} {et~al.}(2016){Ortiz}, {Reffert}, {Trifonov}, {Quirrenbach},
  {Mitchell}, {Nowak}, {Buenzli}, {Zimmerman}, {Bonnefoy}, {Skemer},
  {Defr{\`e}re}, {Lee}, {Fischer}, \& {Hinz}}]{Ortiz2016}
{Ortiz}, M., {Reffert}, S., {Trifonov}, T., {et~al.} 2016, \aap, 595, A55,
  \dodoi{10.1051/0004-6361/201628791}

\bibitem[{{Patience} {et~al.}(2002){Patience}, {White}, {Ghez}, {McCabe},
  {McLean}, {Larkin}, {Prato}, {Kim}, {Lloyd}, {Liu}, {Graham}, {Macintosh},
  {Gavel}, {Max}, {Bauman}, {Olivier}, {Wizinowich}, \& {Acton}}]{Patience2002}
{Patience}, J., {White}, R.~J., {Ghez}, A.~M., {et~al.} 2002, \apj, 581, 654,
  \dodoi{10.1086/342982}

\bibitem[{{Pichardo} {et~al.}(2005){Pichardo}, {Sparke}, \&
  {Aguilar}}]{Pichardo2005}
{Pichardo}, B., {Sparke}, L.~S., \& {Aguilar}, L.~A. 2005, \mnras, 359, 521,
  \dodoi{10.1111/j.1365-2966.2005.08905.x}

\bibitem[{{Queloz} {et~al.}(2000){Queloz}, {Mayor}, {Weber}, {Bl{\'e}cha},
  {Burnet}, {Confino}, {Naef}, {Pepe}, {Santos}, \& {Udry}}]{Queloz2000}
{Queloz}, D., {Mayor}, M., {Weber}, L., {et~al.} 2000, \aap, 354, 99

\bibitem[{{Rafikov}(2005)}]{Rafikov2005}
{Rafikov}, R.~R. 2005, \apjl, 621, L69, \dodoi{10.1086/428899}

\bibitem[{{Raghavan} {et~al.}(2006){Raghavan}, {Henry}, {Mason}, {Subasavage},
  {Jao}, {Beaulieu}, \& {Hambly}}]{Raghavan2006}
{Raghavan}, D., {Henry}, T.~J., {Mason}, B.~D., {et~al.} 2006, \apj, 646, 523,
  \dodoi{10.1086/504823}

\bibitem[{{Raghavan} {et~al.}(2010){Raghavan}, {McAlister}, {Henry}, {Latham},
  {Marcy}, {Mason}, {Gies}, {White}, \& {ten Brummelaar}}]{Raghavan2010}
{Raghavan}, D., {McAlister}, H.~A., {Henry}, T.~J., {et~al.} 2010, \apjs, 190,
  1, \dodoi{10.1088/0067-0049/190/1/1}

\bibitem[{{Reyl{\'e}}(2018)}]{Reyle2018}
{Reyl{\'e}}, C. 2018, \aap, 619, L8, \dodoi{10.1051/0004-6361/201834082}

\bibitem[{{Roell} {et~al.}(2012){Roell}, {Neuh{\"a}user}, {Seifahrt}, \&
  {Mugrauer}}]{Roell2012}
{Roell}, T., {Neuh{\"a}user}, R., {Seifahrt}, A., \& {Mugrauer}, M. 2012, \aap,
  542, A92, \dodoi{10.1051/0004-6361/201118051}

\bibitem[{{Santos} {et~al.}(2017){Santos}, {Adibekyan}, {Figueira},
  {Andreasen}, {Barros}, {Delgado-Mena}, {Demangeon}, {Faria}, {Oshagh},
  {Sousa}, {Viana}, \& {Ferreira}}]{Santos2017}
{Santos}, N.~C., {Adibekyan}, V., {Figueira}, P., {et~al.} 2017, \aap, 603,
  A30, \dodoi{10.1051/0004-6361/201730761}

\bibitem[{{Schlaufman}(2018)}]{Schlaufman2018}
{Schlaufman}, K.~C. 2018, \apj, 853, 37, \dodoi{10.3847/1538-4357/aa961c}

\bibitem[{{Schneider} {et~al.}(2011){Schneider}, {Dedieu}, {Le Sidaner},
  {Savalle}, \& {Zolotukhin}}]{Schneider2011}
{Schneider}, J., {Dedieu}, C., {Le Sidaner}, P., {Savalle}, R., \&
  {Zolotukhin}, I. 2011, \aap, 532, A79, \dodoi{10.1051/0004-6361/201116713}

\bibitem[{{Schwarz} {et~al.}(2016){Schwarz}, {Funk}, {Zechner}, \&
  {Bazs{\'o}}}]{Schwarz2016}
{Schwarz}, R., {Funk}, B., {Zechner}, R., \& {Bazs{\'o}}, {\'A}. 2016, \mnras,
  460, 3598, \dodoi{10.1093/mnras/stw1218}

\bibitem[{{Southworth} {et~al.}(2020){Southworth}, {Bohn}, {Kenworthy},
  {Ginski}, \& {Mancini}}]{Southworth2020}
{Southworth}, J., {Bohn}, A.~J., {Kenworthy}, M.~A., {Ginski}, C., \&
  {Mancini}, L. 2020, \aap, 635, A74, \dodoi{10.1051/0004-6361/201937334}

\bibitem[{{Stassun} {et~al.}(2018){Stassun}, {Oelkers}, {Pepper}, {Paegert},
  {De Lee}, {Torres}, {Latham}, {Charpinet}, {Dressing}, {Huber}, {Kane},
  {L{\'e}pine}, {Mann}, {Muirhead}, {Rojas-Ayala}, {Silvotti}, {Fleming},
  {Levine}, \& {Plavchan}}]{Stassun2018}
{Stassun}, K.~G., {Oelkers}, R.~J., {Pepper}, J., {et~al.} 2018, \aj, 156, 102,
  \dodoi{10.3847/1538-3881/aad050}

\bibitem[{{Thebault} \& {Haghighipour}(2015)}]{ThebaultHaghighipour2015}
{Thebault}, P., \& {Haghighipour}, N. 2015, {Planet Formation in Binaries},
  309--340, \dodoi{10.1007/978-3-662-45052-9_13}

\bibitem[{{Tokovinin}(2014{\natexlab{a}})}]{Tokovinin2014}
{Tokovinin}, A. 2014{\natexlab{a}}, \aj, 147, 86,
  \dodoi{10.1088/0004-6256/147/4/86}

\bibitem[{{Tokovinin}(2014{\natexlab{b}})}]{Tokovinin2014b}
---. 2014{\natexlab{b}}, \aj, 147, 87, \dodoi{10.1088/0004-6256/147/4/87}

\bibitem[{{Tokovinin}(2018)}]{Tokovinin2018}
---. 2018, \apjs, 235, 6, \dodoi{10.3847/1538-4365/aaa1a5}

\bibitem[{{Tokovinin} \& {L{\'e}pine}(2012)}]{Tokovinin2012}
{Tokovinin}, A., \& {L{\'e}pine}, S. 2012, \aj, 144, 102,
  \dodoi{10.1088/0004-6256/144/4/102}

\bibitem[{{Tokovinin} {et~al.}(2006){Tokovinin}, {Thomas}, {Sterzik}, \&
  {Udry}}]{Tokovinin2006}
{Tokovinin}, A., {Thomas}, S., {Sterzik}, M., \& {Udry}, S. 2006, \aap, 450,
  681, \dodoi{10.1051/0004-6361:20054427}

\bibitem[{{Udry} {et~al.}(2004){Udry}, {Eggenberger}, {Beuzit}, {Lagrange},
  {Mayor}, \& {Chauvin}}]{Udry2004}
{Udry}, S., {Eggenberger}, A., {Beuzit}, J.~L., {et~al.} 2004, in Revista
  Mexicana de Astronomia y Astrofisica Conference Series, Vol.~21, Revista
  Mexicana de Astronomia y Astrofisica Conference Series, ed. C.~{Allen} \&
  C.~{Scarfe}, 215--216

\bibitem[{{Udry} {et~al.}(2003){Udry}, {Mayor}, \& {Santos}}]{Udry2003}
{Udry}, S., {Mayor}, M., \& {Santos}, N.~C. 2003, \aap, 407, 369,
  \dodoi{10.1051/0004-6361:20030843}

\bibitem[{{Vigan} {et~al.}(2020){Vigan}, {Fontanive}, {Meyer}, {Biller},
  {Bonavita}, {Feldt}, {Desidera}, {Marleau}, {Emsenhuber}, {Galicher}, {Rice},
  {Forgan}, {Mordasini}, {Gratton}, {Le Coroller}, {Maire}, {Cantalloube},
  {Chauvin}, {Cheetham}, {Hagelberg}, {Lagrange}, {Langlois}, {Bonnefoy},
  {Beuzit}, {Boccaletti}, {D'Orazi}, {Delorme}, {Dominik}, {Henning}, {Janson},
  {Lagadec}, {Lazzoni}, {Ligi}, {Menard}, {Mesa}, {Messina}, {Moutou},
  {M{\"u}ller}, {Perrot}, {Samland}, {Schmid}, {Schmidt}, {Sissa}, {Turatto},
  {Udry}, {Zurlo}, {Abe}, {Antichi}, {Asensio-Torres}, {Baruffolo}, {Baudoz},
  {Baudrand}, {Bazzon}, {Blanchard}, {Bohn}, {Brown Sevilla}, {Carbillet},
  {Carle}, {Cascone}, {Charton}, {Claudi}, {Costille}, {De Caprio},
  {Delboulb{\'e}}, {Dohlen}, {Engler}, {Fantinel}, {Feautrier}, {Fusco},
  {Gigan}, {Girard}, {Giro}, {Gisler}, {Gluck}, {Gry}, {Hubin}, {Hugot},
  {Jaquet}, {Kasper}, {Le Mignant}, {Llored}, {Madec}, {Magnard}, {Martinez},
  {Maurel}, {M{\"o}ller-Nilsson}, {Mouillet}, {Moulin}, {Orign{\'e}}, {Pavlov},
  {Perret}, {Petit}, {Pragt}, {Puget}, {Rabou}, {Ramos}, {Rickman}, {Rigal},
  {Rochat}, {Roelfsema}, {Rousset}, {Roux}, {Salasnich}, {Sauvage}, {Sevin},
  {Soenke}, {Stadler}, {Suarez}, {Wahhaj}, {Weber}, \& {Wildi}}]{Vigan2020}
{Vigan}, A., {Fontanive}, C., {Meyer}, M., {et~al.} 2020, arXiv e-prints,
  arXiv:2007.06573.
\newblock \doarXiv{2007.06573}

\bibitem[{{Wang} {et~al.}(2014){Wang}, {Xie}, {Barclay}, \&
  {Fischer}}]{Wang2014}
{Wang}, J., {Xie}, J.-W., {Barclay}, T., \& {Fischer}, D.~A. 2014, \apj, 783,
  4, \dodoi{10.1088/0004-637X/783/1/4}

\bibitem[{{Ward-Duong} {et~al.}(2015){Ward-Duong}, {Patience}, {De Rosa},
  {Bulger}, {Rajan}, {Goodwin}, {Parker}, {McCarthy}, \&
  {Kulesa}}]{Ward-Duong2015}
{Ward-Duong}, K., {Patience}, J., {De Rosa}, R.~J., {et~al.} 2015, \mnras, 449,
  2618, \dodoi{10.1093/mnras/stv384}

\bibitem[{{Wenger} {et~al.}(2000){Wenger}, {Ochsenbein}, {Egret}, {Dubois},
  {Bonnarel}, {Borde}, {Genova}, {Jasniewicz}, {Lalo{\"e}}, {Lesteven}, \&
  {Monier}}]{Wenger2000}
{Wenger}, M., {Ochsenbein}, F., {Egret}, D., {et~al.} 2000, \aaps, 143, 9,
  \dodoi{10.1051/aas:2000332}

\bibitem[{{White} \& {Ghez}(2001)}]{WhiteGhez2001}
{White}, R.~J., \& {Ghez}, A.~M. 2001, \apj, 556, 265, \dodoi{10.1086/321542}

\bibitem[{{Winn} {et~al.}(2010){Winn}, {Fabrycky}, {Albrecht}, \&
  {Johnson}}]{Winn2010}
{Winn}, J.~N., {Fabrycky}, D., {Albrecht}, S., \& {Johnson}, J.~A. 2010, \apjl,
  718, L145, \dodoi{10.1088/2041-8205/718/2/L145}

\bibitem[{{Winters} {et~al.}(2019){Winters}, {Henry}, {Jao}, {Subasavage},
  {Chatelain}, {Slatten}, {Riedel}, {Silverstein}, \& {Payne}}]{Winters2019}
{Winters}, J.~G., {Henry}, T.~J., {Jao}, W.-C., {et~al.} 2019, \aj, 157, 216,
  \dodoi{10.3847/1538-3881/ab05dc}

\bibitem[{{W{\"o}llert} {et~al.}(2015){W{\"o}llert}, {Brandner}, {Bergfors}, \&
  {Henning}}]{Wollert2015}
{W{\"o}llert}, M., {Brandner}, W., {Bergfors}, C., \& {Henning}, T. 2015, \aap,
  575, A23, \dodoi{10.1051/0004-6361/201424091}

\bibitem[{{Ziegler} {et~al.}(2018){Ziegler}, {Law}, {Baranec}, {Howard},
  {Morton}, {Riddle}, {Duev}, {Salama}, {Jensen-Clem}, \&
  {Kulkarni}}]{Ziegler2018}
{Ziegler}, C., {Law}, N.~M., {Baranec}, C., {et~al.} 2018, \aj, 156, 83,
  \dodoi{10.3847/1538-3881/aace59}

\bibitem[{{Zucker} \& {Mazeh}(2002)}]{ZuckerMazeh2002}
{Zucker}, S., \& {Mazeh}, T. 2002, \apjl, 568, L113, \dodoi{10.1086/340373}

\end{thebibliography}
\bibliographystyle{aasjournal}

\end{document}